%% file: main.tex
\documentclass[journal,onecolumn,draftcls]{IEEEtran}

\usepackage{cite}
\usepackage{amsmath,amssymb,amsfonts}
\usepackage{algorithmic}
\usepackage{graphicx}
\usepackage{textcomp}
\usepackage[mathscr]{eucal}
\usepackage{graphicx}
\usepackage{latexsym}
\usepackage{amssymb}
\usepackage{psfrag}
\usepackage{epsfig}
\usepackage{MnSymbol}
\usepackage{enumerate}
\usepackage{multirow}
\DeclareMathAlphabet{\mathpzc}{OT1}{pzc}{m}{it}

\usepackage{amsfonts}
\usepackage{graphicx}
\usepackage{amsfonts}
\usepackage{color}
\usepackage[textsize=small]{todonotes}
\usepackage{caption,subcaption}
\usepackage{soul}

\usepackage{setspace}
\newcommand{\one}{{\boldsymbol{1}}}

\newcommand{\E}{\mathbb{E}}

\newcommand{\PP}{\mathbb{P}}

\newcommand{\ti}{\tilde}

\newcommand{\beq}{\begin{eqnarray*}}
\newcommand{\eeq}{\end{eqnarray*}}
\newcommand{\beqn}{\begin{eqnarray}}
\newcommand{\eeqn}{\end{eqnarray}}

\newcommand{\bt}{\begin{theorem}}
\newcommand{\et}{\end{theorem}}

\newcommand{\be}{\begin{equation}}
\newcommand{\ee}{\end{equation}}

\newcommand{\NN}{\mathbb{N}}

\newcommand{\N}{\mathbb{N}}

\begin{document}

\title{Generalization of LRU Cache Replacement Policy with Applications to Video Streaming}
\date{}

\author{Eric Friedlander and Vaneet Aggarwal \thanks{E. Friedlander is with the University of North Carolina at Chapel Hill, Chapel Hill, NC 27599, USA, email: ebf2@email.unc.edu. V. Aggarwal is with  Purdue University, West Lafayette, IN 47907, USA, email: vaneet@purdue.edu}}

\maketitle

\input{abstract}

\input{intro}

\input{related}

\input{model}

\input{result}

\input{numerical}

\input{conclusion}

\input{apdx}

\bibliographystyle{IEEEtran}
\bibliography{references}

\end{document}

%% file: abstract.tex
\begin{abstract}
Caching plays a crucial role in networking systems to reduce the load on the network and is commonly employed by content delivery networks (CDNs) in order to improve performance.
One of the commonly used mechanisms, Least Recently Used (LRU), works well for identical file sizes. However, for asymmetric file sizes, the performance deteriorates. This paper proposes an adaptation to the LRU strategy, called gLRU, where the file is sub-divided into equal-sized chunks. In this strategy, a chunk of the newly requested file is added in the cache, and a chunk of the least-recently-used file is removed from the cache. Even though approximate analysis for the hit rate has been studied for LRU, the analysis does not extend to gLRU since the metric of interest is no longer the hit rate as the cache has partial files. This paper provides a novel approximation analysis for this policy where the cache may have partial file contents. The approximation approach is validated by simulations. Further, gLRU outperforms the LRU strategy for a Zipf file popularity distribution and censored Pareto file size distribution for the file download times. Video streaming applications can further use the partial cache contents to help the stall duration significantly, and the numerical results indicate significant improvements (32\%) in stall duration using the gLRU strategy as compared to the LRU strategy. 
Furthermore, the gLRU replacement policy compares favorably to two other cache replacement policies when simulated on MSR Cambridge Traces obtained from the SNIA IOTTA repository.
\end{abstract}

\IEEEkeywords{ Caching, Least Recently Used, Video Streaming, Characteristic Time Approximation, Che's approximation, }

%% file: intro.tex
\section{Introduction}\label{sec:intro}
In order to improve the performance of web-based services (e.g., cloud-based storage systems, Video-on-Demand (VoD), etc.), content delivery architectures frequently employ a caching system.
A typical system consists of a set of large centralized servers storing a large set of documents (e.g. videos)
and a network of distributed servers (caches).
The caches are closer to the user, and thus allow for faster download speeds.
However, since the caches are small relative to the size of the centralized servers, one needs to employ a set of rules governing which files are stored on each cache, referred to as a ``Cache Replacement Policy''.

One of the most popular policies is the so-called ``Least Recently Used'' (LRU) replacement policy.
Under the LRU policy, the cache can be thought of as a queue.
When a new file is requested, it is added to the head of the queue (or moved to the head if it is already present in the queue).
If a file reaches the tail of the queue it is ``pushed out'' (i.e., removed from the cache).
Since the most popular files are the ones which are requested most often, they have a much higher probability of being stored in the cache resulting in faster content delivery.
As a result, the performance and properties (both theoretical and empirical) are a topic of much research (cf. \cite{breslau1999web,fagin1977asymptotic,che2002hierarchical,dan1990approximate,dernbach2016cache,fricker2012versatile,
fricker2012impact,jelenkovic2008characterizing,mahanti2000traffic,rosensweig2010approximate,gallo2014performance,
zhang1999web,Rosensweig2013} and references therein).

There are many extensions of the classical LRU policy, including q-LRU and LRU(m) \cite{garetto2016unified, gast2016asymptotically}, that mainly focus on situations where entire files are cached.
One of the issues with the LRU policy is that a large file request will evict multiple small files in the cache and thus can hurt the system performance. In order to avoid this effect, this paper  proposes a generalization of  the LRU replacement policy (denoted as gLRU). In general, files can be divided into equally sized pieces or chunks of equal size.
The gLRU replacement policy differs from LRU in that, when a file is requested, only one additional chunk is added to the cache (unless all chunks of the file exist already in the cache). 
For example, suppose a document with 100 chunks is requested and 10 chunks are currently cached.
Under gLRU, the 10 chunks will be moved to the head of the cache along with 1 additional chunk.
In the LRU policy, the entire file will be added.
We assert that gLRU results in improved performance (i.e., faster download speeds, fewer delays, etc.) than the original LRU. 

Even though gLRU is a general approach and for identical sized files with a single chunk reduces to the LRU policy, analysis of the gLRU scheme is an important and challenging problem. 
As with LRU, the computation of hit rates, the probability that a file is in the cache in the steady state, cannot be characterized easily in closed form.
There have been multiple approximation methods to characterize the approximate hit rate for LRU, with one of the commonly known approximations called the ``characteristic time approximation'' \cite{fagin1977asymptotic, che2002hierarchical}. With the gLRU caching policy, a file is partly in the cache. Thus, the metric of interest is not the hit rate, but the distribution of the number of chunks of each file in the cache in the steady state. Thus, the analysis of gLRU brings a new dimension in solving the problem and the previous analysis which adds or removes the entire file in the cache cannot be readily applied. The main result of the paper is the approximation for the steady state distribution for the gLRU strategy. The proposed approximation is also validated through simulation. Even though multiple approximation techniques have been proposed for cache replacement strategies, this is the first work to the best of our knowledge that analyzes the distribution of files in cache, where partial files can exist in the cache and the caching strategies adds/removes a chunk of the file rather than the complete file.
The method proposed in \cite{wu2001segment} considers a similar policy targeted toward a video streaming setting in which portions of files can be stored, however no present analytical results characterizing the number of cached files in steady state are given.

This paper further aims to compare the performance of gLRU with  LRU and give preliminary comparisons with several other cache replacement policies.
In addition to LRU, we consider two alternative policies. We now give a brief description of these two policies with a more in-depth description given in Section \ref{sec:AlternativePolicies}.
The first alternative, outlined in \cite{wu2001segment}, aims to improve the performance of video streaming by first grouping the chunks into segments such that the $i$-th segment contains $2^{i-1}$ chunks and we refer to this policy as \emph{segLRU}.
This segmentation of the chunks means that as a video is requested more frequently, the number of cached chunks increases faster than with a simple LRU policy.
The policy then divides the cache into two sub-caches, the first for segments consisting of chunks occurring early on in the video and the second for later segments.
The two sub-caches operate according to a modified LRU policy.
The performance improvement of this policy lies in the fact that early portions of requested videos are more likely to be stored since there is a cache which is devoted to these segments.
We also consider the AdaptSize method proposed in \cite{berger2017adaptsize}.
AdaptSize is a probabilistic caching policy in which smaller files are stored with higher probabilities, and these probabilities are dynamically tuned to improve performance.
In order to compare these methods, multiple performance metrics are considered, including the proportion of chunks retrieved from the cache (as a generalization to the average hit rate), and the download time of the file. 
Further, video specific metrics like the stall duration are also considered. 

We observe for a Zipf file popularity distribution and censored Pareto file size distribution that the gLRU cache policy outperforms the LRU policy in all the metrics.
These results indicate that the flexibility of adding and removing a chunk of the file in contrast to the entire file can improve system performance, at least in a setting with static popularity distribution.
When compared with segLRU and AdaptSize on production traces the results are also promising (cf. Figures \ref{tab:gvseg} and \ref{tab:gvas}).
Specifically, gLRU outperforms segLRU in between 43-46 out of 48 trials (depending on the performance metric considered). Furthermore, the median improvement in performance ranges from 1.79\% to 78.2\% (depending on the metric).
When compared to AdaptSize, gLRU performs better in all 48 trials for the cache miss rate $p_m$ and download time $T_w$ .
In addition it outperforms AdaptSize in 27, 45, and 36 out of 48 trials for the proportion of chunks retrieved from the cache $p_c$, the delay time $T_d$, and the proportion delayed $p_d$, respectively.
The median improvement for each metric ranges from .16\% for $p_c$ to 96.1\% for $T_w$.
The proposed scheme borrows the advantages of LRU, including  low complexity, ease of implementation, and being adaptive to the change in the arrival distributions. In addition, gLRU outperforms LRU by not having to displace multiple small files to add a larger file. For VoD settings, the availability of earlier chunks in the cache can further help since the later chunks have later deadlines. The proposed scheme will have partial files in the cache, and that can help the stall duration since only the later chunks must be retrieved. Numerical results indicate a median improvement of 20\%, 31\%, and 32\% in proportion of chunks accessed from cache, download time, and video stall duration, respectively, using the gLRU caching strategy as compared to the LRU caching strategy. 
It is important to note that we only consider a static popularity distribution and gLRU may suffer when the popularity distribution is rapidly changing.
To combat this, one can consider extensions of gLRU in which more and more chunks are added each time a file is requested in order to improve the policies reactivity (cf. Section \ref{sec:repPol}).
However, we leave a more in-depth analysis of gLRU's reactivity to future work. 
Another topic of concern is the dependence structure between the file size and popularity distributions.
While the file size and popularity distributions are assumed independent in the synthetic traces, we note that the joint distribution is certainly important for the performance of gLRU.
In Appendices \ref{sec:dependence} and \ref{app:corrhist}, we explore the effect of positive and negative correlation between the the file size and popularity distributions.
In the case of negative correlation, the more popular files are smaller and thus more of them can be stored in the cache leading to improved performance (cf. Table \ref{tab:posvneg} and Figure \ref{fig:histcomp}).
Furthermore, gLRU outperforms LRU in both the positive and negative correlation cases but the effects are more pronounced in the case of negative correlation (cf. Table \ref{tab:posneg}).

The rest of the paper is organized as follows. Section \ref{related} reviews some of the related work in the analysis of LRU caching, as well as some proposed variants of LRU to deal with the issue of asymmetric file sizes.  In Section \ref{sec:model} we describe the model which is the subject of our analysis including a description of our proposed cache replacement policy, gLRU.
Section \ref{sec:newche} then gives the proposed approximation to the  gLRU caching policy.
In Section \ref{sec:LRUcompare}, we use both the original characteristic time approximation and the  approximations proposed in Section \ref{sec:newche} to compare LRU with gLRU.
In particular, we show that since gLRU does not need to add entire files to the cache, it is able to store smaller pieces of more files in the cache.
Sections \ref{sec:chevalid} and \ref{sec:numeval} are devoted to the results of numerical simulations. 
In Section \ref{sec:chevalid}, we demonstrate that the proposed  approximation to gLRU is valid and in Section \ref{sec:numeval}, evaluation of gLRU against several alternative policies are given on both synthetic and production trace data.
Finally, Section \ref{sec:conclusion} concludes this paper.

%% file: related.tex
\section{Related Work}\label{related}

{\bf Analysis of LRU Caching: } Evaluating the performance of cache networks is hard, considering that the Markov chain associated with  a single LRU (Least Recently Used) cache has an exponential number of states  \cite{king1972analysis,flajolet1987birthday,dan1990approximate}.
Multiple approximation approaches have been studied \cite{breslau1999web,fagin1977asymptotic, che2002hierarchical,dan1990approximate,dernbach2016cache,fricker2012versatile,	fricker2012impact,jelenkovic2008characterizing,mahanti2000traffic,rosensweig2010approximate,gallo2014performance,
	zhang1999web,Rosensweig2013} in the literature, with two key types of analysis techniques - the characteristic time approximation  \cite{flajolet1987birthday,fagin1977asymptotic, che2002hierarchical} and the network calculus approach \cite{Rosensweig2013}. One of the key metrics to quantify in caching systems is the hit rate, which describes the probability of finding a file in the cache given a popularity distribution on the set of available content.  The authors of \cite{flajolet1987birthday} provided an expression for hit rate of identical-sized files, which can be computed using numerical integration approaches. The authors of \cite{fagin1977asymptotic, che2002hierarchical} presented a method for  approximating (called the characteristic time approximation) the hit rates for such a system assuming that all files are of identical size. However,  in most cases files are of different sizes. Further work by the authors of \cite{fricker2012versatile} provide the theoretical machinery behind the efficacy of the characteristic time approximation  and provide a simple extension to the case of multiple file sizes (cf. equation (5) therein).

{\bf Adaptations of LRU Caching: } 
Several extensions of LRU have been proposed.
\cite{garetto2016unified} outlines a variety of these methods, specifically q-LRU, k-LRU, RANDOM, and k-RANDOM.
The q-LRU policy is the same as LRU except that files are only added with probability $q$. 
An extension of q-LRU called $q_i-$LRU in which each file has its own probability of being added to the queue is presented in \cite{neglia2017access}. This probability is computed based on the content size and time required to retrieve the content from the cache .
In k-LRU, requested files must traverse $k-1$ additional virtual LRU caches before it is added to the actual cache.
RANDOM and k-RANDOM are the same as q-LRU and k-LRU except files are evicted from the cache at random.
Another method, called LRU(\textbf{m}), is outlined in \cite{gast2015transient,aven1987stochastic}.
This policy exploits $h$ caches of sizes $m_1,\ldots,m_h$ in which the first $v$ caches are virtual and of sizes $m_1,\ldots,m_v$ while the remainder are real. For more details see \cite{gast2016asymptotically} which gives an analysis of both k-LRU and LRU(m).
Several other probabilistic replacement policies are proposed in \cite{starobinski2001probabilistic}, where the authors consider files which have non-uniform size and non-uniform access costs.
However, the setting of \cite{starobinski2001probabilistic} differs from ours in that it does not consider chunked files.
We note that in future work, it maybe possible to combine the above policies with gLRU by, for example, considering a sequence of caches employing a gLRU replacement policy instead of a pure LRU policy.

 One of the key issues in the LRU based caching strategy is that a large file arrival can evict multiple small files. 
 In order to have better performance with realistic file sizes, multiple approaches have been proposed, see \cite{Cidon16,berger2017adaptsize} and the references therein. An admission control strategy in \cite{berger2017adaptsize} is used to decrease the probability that a large file size is added in the cache. One of the potential issues with these strategies is that the  addition and removal from the cache is still at the file level. 
 In a VoD setting, one may achieve better performance from simply caching the early portions of each file as in \cite{wu2001segment}. The time it takes to watch the early portion of the video provides a buffer during which the later segments of the video can be returned from a centralized server.
 In contrast, this paper provides the flexibility of adding/removing files at the chunk level. The authors of \cite{Sprout} considered the flexibility for adding/removing chunks from the cache, however, the caching strategy is based on knowing the file arrival rates.  
 Even though a window-based scheme can be used to learn the arrival rate and use such caching policies, the complexity for the cache update is higher. In contrast, this paper uses a generalization of the LRU caching strategy that is completely adaptive to the arrival pattern, does not take any such parameters in the strategy, and is easy to implement. 
Similarly, in \cite{jin2000popularity}, the author's propose a replacement policy based on the Greedy Dual-Size algorithm which leverages information based on the file popularities which are estimated online.
However, despite considering varying file sizes, the authors do not consider chunked files.

{\bf Cache Management for Video Streaming:}
There is a large amount of literature on the particulars of video caching and, in some cases, its relationship with the LRU caching policy.
Many of these are concerned with the intricacies of how to design hierarchical network architectures to improve system performance.
This differs from the work presented here in that our proposed cache replacement is considered on a more granular level, considering the performance on a single cache.
For example, four different caching algorithms for planet-scale CDNs called xLRU, Cafe Cache, Optimal Cache, and Psychic Cache are outlined in \cite{mokhtarian2014caching}.
These methods are concerned with networking together large sets of servers and a large portion of their work is devoted to evaluating when to redirect traffic between servers versus when to add files to caches at which a requests arrive.
Another paper is \cite{rossini2014multi} in which the authors consider a two-layered architecture tailored to video streaming for chunked files.
In the presented system, the second level caches employ an LRU caching policy and the model exploits correlations between files chunks of the same video in order to prefetch later file chunks in order improve the performance of video streaming.
It is an interesting topic of future work as to whether some of these network architectures can be improved by incorporating gLRU instead of LRU caches.
Another work of interest is \cite{wu2001segment}, in which the authors also consider chunked video files.
In this work, the chunks are divided into segments containing multiple chronologically adjacent chunks.
Chunks from later on in a given video are combined into larger and larger segments.
Furthermore, the cache is divided in two. 
The first portion is devoted to smaller, early (i.e., those segments corresponding to the beginning of a video) segments operating under an simple LRU policy.
The second cache handles the larger, later segments and employs a probabilistic LRU policy in which the relative popularity and size of each segment is used to determine its admission probability.
Throughout this work we will refer to this policy as segLRU.
In Sections  \ref{sec:videoStream} and \ref{sec:trace}, we compare gLRU with segLRU and the AdaptSize method outlined earlier.

%% file: model.tex
\section{System Model and Problem Formulation }\label{sec:model}
In this section, we outline the  system  model  and describe the caching policy that will be analyzed in this paper.

\subsection{System Parameters }\label{sec:netmodel}
Consider a single server storing a collection of $N$ files. 
The files may be of different sizes and file $i\in\{1,\ldots,N\}$ is comprised of $s(i)$ equally sized pieces, or chunks. 
While the number of chunks which comprise each file may vary, the size of each individual chunk is the same for all files.
In video streaming settings, each chunk typically consists of 1-4 seconds of video \cite{liu2008substream}. Without loss of generality, we assume all chunks have unit size.

We assume a single user in the system, which may be an aggregate point for multiple users (e.g., edge router).
The user requests the files from the server, where the aggregate arrival process follows a Poisson process with rate $\lambda$. The probability that a given request is for file $i\in\{1,\ldots,N\}$ is proportional to its popularity $q(i)$. In general, $q$ is assumed to follow a Zipf law with parameter $\alpha$ (cf. \cite{fagin1977asymptotic,fricker2012versatile,che2002hierarchical,zhang1999web} and references therein for a more in-depth discussion).  Without loss of generality, we assume that $\lambda = \sum_{i=1}^Nq(i)$ and the arrival rate for file $i$ is $q(i)$. This implies that the probability that a given request is for file $i$ is $q(i)/\lambda$.

When a file request arrives at the server, all chunks are entered in a first-in-first-out (FIFO) queue which services the requests. We assume that distribution of service times (from the server) for any chunk is exponential with a rate $\mu$.
In order to improve retrieval speed, the system employs a cache of capacity $C$ in which any file chunk requires one unit of capacity. 
For approximating the steady state hit probabilities in Section \ref{sec:newche}, we assume that the processing rate $\mu$ is infinite.
This assumption is in line with \cite{fagin1977asymptotic,fricker2012versatile,che2002hierarchical} in which it is assumed that files are cached instantaneously upon request.
For the numerical evaluation of the gLRU policy in Section \ref{sec:numeval}, we assume a finite $\mu$.
Furthermore, the steady state hit probabilities can be used to compute performance metrics (e.g. download time) for a system with finite $\mu$.
A file $i$ finishes downloading when all the $s(i)$ chunks of the file are available to the user.  However, when considering video streaming, a user can begin playing the video before all the chunks are received. More about this will be discussed in Section \ref{sec:videoStream}.
One can think about the cache as being ``close'' to the user and having a service time which is negligible. In the next subsection, we will describe the cache replacement policy that will be used to decide the contents in the user cache.

\subsection{Cache Replacement Policy}\label{sec:repPol}
In order to achieve a better user experience (e.g., higher latency, less buffering, etc.), one can consider several different policies for allocating space within the cache.
One commonly used policy is the Least Recently Used (LRU) replacement policy \cite{fagin1977asymptotic,breslau1999web,che2002hierarchical,dan1990approximate}.
When employing the LRU replacement policy, all pieces of a requested file are moved to the head of the cache.
If there is not enough capacity to accommodate these new file requests, the files at the tail of the cache (i.e., those which were the least recently requested) are forced out. One of the key issues with LRU is that different files have different sizes, and a large file-size request can displace multiple files with lower file size.  In this work, we propose a generalization of LRU which we will refer to as gLRU. 
In the gLRU replacement policy, instead of adding all file chunks to the head of the cache, only the pieces already in the cache plus one additional chunk (should one exist) are added, thus increasing the number of cached pieces of the requested file by at most one. The proposed gLRU scheme is easy to implement since the existing chunks of the file are moved to the head with one additional chunk and one chunk at the tail is removed.
 We note that the LRU policy can be implemented using concurrent linked hashed maps \cite{surtani2014concurrent,codehash}. This can be easily modified to add the counter of chunks in the cache. The changes for gLRU include decreasing the number of chunks rather than removal of the file index at the end of the doubly linked list (unless there is only one chunk already), and adding a chunk when the file is moved to the head. Thus, gLRU can be implemented using concurrent linked hashed maps with minor adaptations on the LRU implementation.
 Further, this scheme is online and adapts to the changing file arrival distribution.  
As we will show in Section \ref{sec:videoStream}, numerical results demonstrate that gLRU has superior performance on many performance metrics of interest, including file download times and video stall duration.

We note that adding chunks one-by-one may be slow in adapting and has the potential to yield poor performance when the popularity distribution is rapidly changing.
In order to improve the reactivity of gLRU, one could consider simple extensions of the algorithm outlined here.
For example, by dividing the chunks into segments as in \cite{wu2001segment} so that each time the file is requested a larger portion is added.
These extensions fit quite simply within the approximation heuristic framework proposed in Section \ref{sec:newche}, and equivalent approximations of hit probabilities can be derived using the same steps as outlined in this work.

With this in mind, we propose a further generalization of LRU which we refer to as gLRU($d$) inspired by the transport control protocol (TCP)  congestion control algorithm \cite{hoe1996improving,liu2019optimizing}. 
In particular, in TCP CUBIC, the  window size is a cubic function of time since the last congestion event \cite{ha2008cubic}. 
Since the window size increases as a cubic function rather than increasing by $1$ as in TCP Reno, TCP cubic has been shown to produce less bursty traffic \cite{simon2017large}.    In gLRU($d$), when a file is requested, instead of adding one additional chunk, an additional $p^d$ chunks are added where $p$ is the number of times that file has been requested since it last dropped out of the cache (To draw a parallel with the TCP congestion control algorithm, the last time it dropped out of cache is equivalent to the last congestion event and the amount of chunks in the cache is equivalent to the window size).
We can tune the reactivity of the replacement policy by changing the value of $d$ in that increasing $d$ will increase the reactivity of the algorithm.
As previously mentioned, the approximation heuristic proposed in Section \ref{sec:newche} can easily be extended to gLRU($d$).
However, as we have only numerically validated the approximation for gLRU  (i.e. gLRU(1)) we present the heuristic in Appendix \ref{sec:newched}.
We leave a further detailed analysis of the effects on the reactivity and the question of how to optimally choose $d$ for future work.

\subsection{Problem Formulation}

Fagin first  proposed a simple approach for estimating the hit rates of a cache operating under the LRU strategy \cite{fagin1977asymptotic}. The method was then rediscovered in \cite{jelenkovic2004optimizing,che2002hierarchical}. We call this approach the characteristic time approximation approach and note that other papers in the area \cite{fricker2012versatile,garetto2016unified} use the term Che's approximation, even though the paper \cite{che2002hierarchical} has multiple co-authors. In this paper, we aim to estimate the probability distribution of the number of chunks of a file in the cache in the steady state using the gLRU caching policy. Further, this paper aims to see improvement of the proposed caching strategy as compared to the LRU strategy.

%% file: result.tex
\section{Generalization of the Characteristic Time Approximation}

In this section, we will provide an approximation for the distribution of cache contents using the  gLRU caching strategy and discuss the results with a comparison to the LRU caching policy.

\subsection{Characteristic Time Approximation for LRU Caching Policy}
In this subsection, we will describe the key approach in the characteristic time approximation \cite{fagin1977asymptotic, che2002hierarchical}. The characteristic time approximation  gives a simple, efficient, and accurate method of estimating the hit probabilities (i.e., the probability of finding a given file in the cache) for a system employing the LRU replacement policy. While the approximation was established for a system in which all files had one chunk of equal size, the method can easily be extended to files of multiple sizes (c.f. \cite{fricker2012versatile}). Let $\ti\tau_{i,1}$ represent the time of the first request for file $i$ (the first inter-arrival time). The characteristic time approximation, applied to files of multiple sizes (c.f. (5) of \cite{fricker2012versatile}) relies on defining, for each file $n\in\{1,\ldots,\N\}$, random variables
\begin{align}\label{eqn:Xdef}
X_n(t) = \sum_{i = 1,i\neq n}^N\one_{\{\ti\tau_{i,1}<t\}}s(i)
\end{align}
and
\begin{align*}
T_C(n) = \inf\{t>0:X_n(t) \geq C\}
\end{align*}
where $X_n(t)$ represents the number of file chunks, other than file $n$, that will be added to the head of the cache by time $t$.
Assuming that object $n$ is inserted into the cache at time 0, it follows that $T_C(n)$ represents the amount of time it will take for at least $C$ chunks to be added to the cache at which point all pieces of file $n$ will have fallen out of the cache if $n$ has not been requested.

Ultimately, $T_C(n)$ is estimated by setting $C$ equal to the expected value of $X_n$,
\begin{align}\label{eqn:exp1}
\E X_n(T_C(n)) = \E\sum_{i \neq n}(1-e^{-q(i)T_C(n)})s(i)
\end{align}
however, some assumptions need to made.
The first is that the cache is sufficiently large as to assume that $T_C(n)$ is deterministic.
The second is to assume that $T_C(n) = t_C$ is the same for all $n$ where $t_C$ solves
\begin{align}\label{eqn:origche1}
C = \sum_{i = 1}^N(1-e^{-q(i)t_C})s(i).
\end{align}
It is also assumed that it is sufficient to consider,
\begin{align*}
X(t) = \sum_{i = 1}^N\one_{\{\ti\tau_{i,1}<t\}}s(i)
\end{align*}
to estimate $t_C$ rather than each $X_n$ and thus \eqref{eqn:exp1} becomes
\begin{align}\label{eqn:origche2}
C = \E X(t_C) = \sum_{i \neq 1}^N(1-e^{-q(i)t_C})s(i)
\end{align}
which is the same as \eqref{eqn:origche1}.
Such an assumption is valid if the popularity of any individual file is small compared to the total popularity (i.e. $\sum_{i}q(i)$).
Indeed, this is the case for the Zipf popularity distribution.
An estimate for $t_C$ is then obtained by numerically solving \eqref{eqn:origche2}.

\subsection{Challenges for Analysis of gLRU}

The key difference in gLRU as compared to LRU is that the cache contains partial files. For each file request, at most one chunk is added in gLRU whereas in LRU where the entire file is added. Similarly, gLRU removes a chunk of a file at a time in contrast to the removal of entire file(s) in LRU.

The difficulty in extending the characteristic time approximation to the gLRU caching policy is that the number of chunks added to the head of the cache is a random variable dependent on the state of the system, the file size distribution, and the popularity distribution.
In \cite{fricker2012versatile,che2002hierarchical,fagin1977asymptotic}, the state of the cache at time zero is unimportant and thus one can assume that it is empty.
In the case of gLRU, the number of chunks added to the head of the cache due to a file request is dependent on the current state of the cache.
As a result, it is difficult to write down an exact expression equivalent to \eqref{eqn:Xdef}.

\subsection{Proposed Approximation for gLRU($d$)}\label{sec:newche}

In this subsection, we provide an approximation for the probability of $j$ chunks of file $n$ in the cache when the gLRU caching policy is used.

Since an exact expression equivalent to \eqref{eqn:Xdef} is hard to write, we write an approximation of $X_n(t)$ by replacing $s(i)$ in the $i$-th term in the sum with the expected number of file $i$ chunks in the cache when the system is in steady state.  This expected value can be computed given $t_C$, and we will denote it as $d(i,t_C)$.

Let $\tau_{n,k}$ represent the $k^{\text{th}}$ inter-arrival time of requests for file $n$.
Given a $t_C$, the probability of finding at least $j$ chunks of file $n$ in the cache is equal to the probability that the last $j$ inter-arrival times, looking back from the current time, for file $n$ are less than $t_C$.
Assuming that the probability of a file being requested while it is at the end of the cache is small, this can be approximated as follows,
\begin{align}\label{eqn:hitprob}
h_j(n,t_C) \doteq \PP(\tau_{n,1}<t_C,\ldots, \tau_{n,j}<t_C) = (1-e^{q(n)t_C})^j,
\end{align}
which follows since the $\tau_{n,j}$'s are independent and identically distributed exponential random variables with rate parameter $q(n)$.
We note that this is not really a new assumption but a result of the rate of requests for file $i$ being small compared to that overall rate of requests under a Zipf popularity law.
Without this assumption, the expression for $h_j$ would need to account for all cases in which the specified file is requested after several chunks of the file had fallen out of the cache and $j-1$ were remaining, vastly complicating \eqref{eqn:hitprob}.
Suppose one starts with $j'>j$ chunks in the cache.
If a file request arrives when $j'-j+1$ chunks have fallen out of the end of the cache then $j$ will be added to the head.
The assumption implies that this event can be ignored.
Let $Y_i,i\in\{1,\ldots,N\}$ be random variables such that $Y_i$ represents the number of cached chunks of file $i$ when the system is in steady state.
Recalling that $d(i, t_C)$ denotes the expected number of file $i$ chunks in the cache when the system is in steady state, it then follows that
\begin{align*}
d(i,t_C)
= \sum_{k=1}^{s(i)}\PP(Y_i\geq k)
= \sum_{k=1}^{s(i)}(1-e^{q(i)t_C})^k.
\end{align*}
The proposed approximation of $X_n(t)$ is then given as
\begin{align*}
\ti X_n(t) = \sum_{i = 1,i\neq n}^Nd(i,t).
\end{align*}
As in the derivation of \eqref{eqn:origche2}, it is sufficient to consider
\begin{align*}
\ti X(t) = \sum_{i = 1}^Nd(i,t)
\end{align*}
and one can then estimate $t_C$ by setting the expected value of $\ti X$ to $C$ and solving for $t_C$. 
This amounts to solving the following equation,
\begin{align*}
C = \E\ti X = \sum_{i=1}^N\sum_{k=1}^{s(i)}(1-e^{q(i)t_C})^k.
\end{align*}

Once an estimate for $t_C$ is obtained, one can compute hit probabilities using \eqref{eqn:hitprob}.
For example, the probability of finding at least one piece of file $n$ in the cache is simply $h_1(n,t_C)$, where $h$ is as in \eqref{eqn:hitprob}.
This can then be used to find other metrics of interest.
For example, the probability that exactly $j$ pieces of file $n$ are cached when the system is in steady state is given as
\begin{eqnarray}
&&\PP(\text{at least }j\text{ pieces of file }n\text{ are cached})\nonumber\\
&&\qquad-\PP(\text{at least }j+1\text{ pieces of file }n\text{ are cached})\nonumber\\
&=&h_j(n,t_C) -h_{j+1}(t_C)\nonumber\\
&=& (1-e^{q(n)t_C})^j-(1-e^{q(n)t_C})^{j+1}.\label{jincache}
\end{eqnarray}

\subsection{Discussion}\label{sec:LRUcompare}

One important implication of the assumptions made in the characteristic time approximation is that the  popularity of any given file is small relative to the total popularity.
Indeed, the simulations in \cite{fricker2012versatile,che2002hierarchical} and Section \ref{sec:chevalid} suggest that this is reasonable.
In this work, we further use this assumption when approximating the hit rates via \eqref{eqn:hitprob} to imply that the amount of time a file spends at the end of the cache is small, and thus, the probability a request arrives while  a few (but not all) chunks have fallen out of the cache is negligible. 
When contrasting LRU and gLRU, this implies that the LRU replacement policy results in an ``all or none'' scheme in which files are either entirely in the cache or not at all in the cache while gLRU is able to store portions of a greater number of files.

\begin{figure}[htbp]
	\includegraphics[width = .5\textwidth]{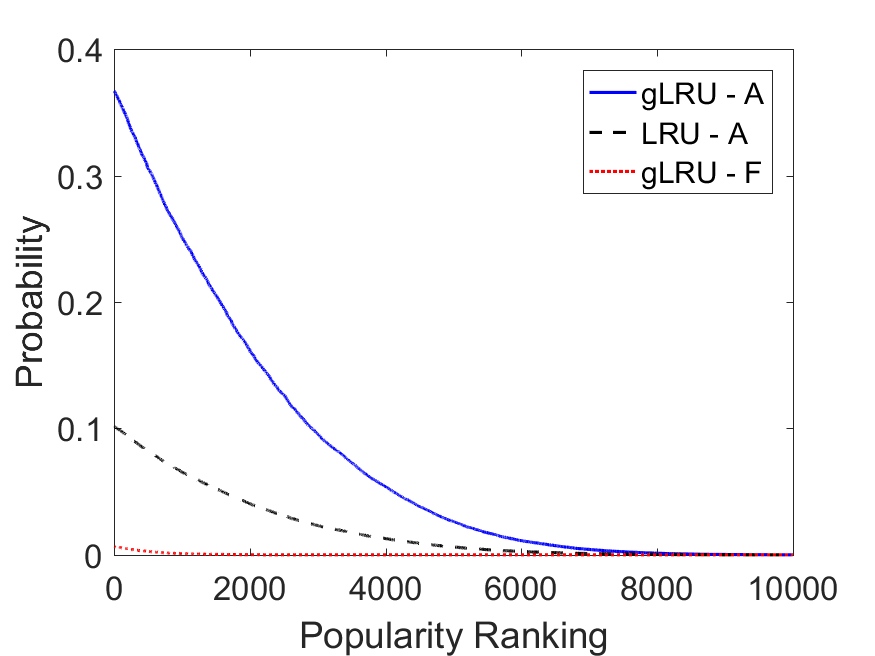}
	\caption{Comparison of hit rates for LRU and gLRU cache replacement policy. gLRU-A and LRU-A lines indicate the probabilities of finding at least one piece of a file of a given popularly (rank is given on $x$-axis) in the cache under the gLRU and LRU replacement policies, respectively. gLRU-F line gives the probability of finding all chunks of the file under the gLRU policy. Note that there is no need for an analogous LRU line since this value doesn't change for LRU since it is an ``all or nothing'' policy.}
	\label{fig:LRUgLRUComp}
\end{figure}

In Figure \ref{fig:LRUgLRUComp}, we present estimates from the characteristic time approximation and the proposed approximation that the cache contains any chunks of a file of given popularity for LRU  and gLRU. Further, Figure \ref{fig:LRUgLRUComp} also depicts the probability of finding all chunks of a file (i.e. the entire file) under the gLRU replacement policy. Since the LRU policy stores either all or no chunks of each file, an analogous line is not needed as it would coincide with the earlier LRU line. 
In this particular example $\alpha = .8$, all files have 5 chunks, and the cache can hold 10,000 chunks at a time.
Similar results can be obtained by varying $\alpha$ or the number of chunks.
If we allow the number of chunks to be random, similar patterns are obtained for the probabilities of observing any chunks in the cache, however the probabilities of finding full files in the cache under gLRU becomes dependent on the individual file sizes.

In the Video-on-Demand (VoD) setting, the user can begin watching as soon as a single chunk has been retrieved.
The time it takes to view this chunk (and the other chunks stored in the cache) provides a cushion during which portions of the video which occur later, and are not stored in the cache, can be retrieved from the centralized server.
For this reason, it makes sense that gLRU should be superior in the VoD setting.
In Section \ref{sec:videoStream}, we provide numerical results showing that, in addition to this improvement in delay times, total download time is also reduced by using gLRU. 
The flexibility of having partial file chunks can indeed help getting the number of file chunks in the cache roughly proportional to the arrival rates, helping to improve the file download times.

%% file: numerical.tex
\section{Numerical Validation of the Proposed Approximation}\label{sec:chevalid}
In this section, we present the result of several numerical simulations showing the validity of the approximations established in Section \ref{sec:newche}.
We validate the approximations presented in Section \ref{sec:newche} via simulation. 
The simulated system contains $N=10,000$ different files with popularity law $q(i)\sim\text{Zipf}(\alpha)$. Each file has a constant number of chunks $c$. The reason for this is to separate the performance differences due to file size and popularity. If one were to consider distributions on popularity and file size then one would need to account for the joint distribution between the two, an interesting question on its own (see Section \ref{sec:dependence}). 
While not presented here, we did consider Pareto file sizes and the results were  qualitatively similar. Pareto file sizes have been used in Section \ref{sec:videoStream} to compare performance of LRU and gLRU.
The arrivals of requests for file $i\in\{1,\ldots,N\}$ is a Poisson process with rate $q(i)$. 

\begin{figure*}
\centering
\begin{subfigure}[t]{0.33\textwidth}
\includegraphics[width = 1\textwidth]{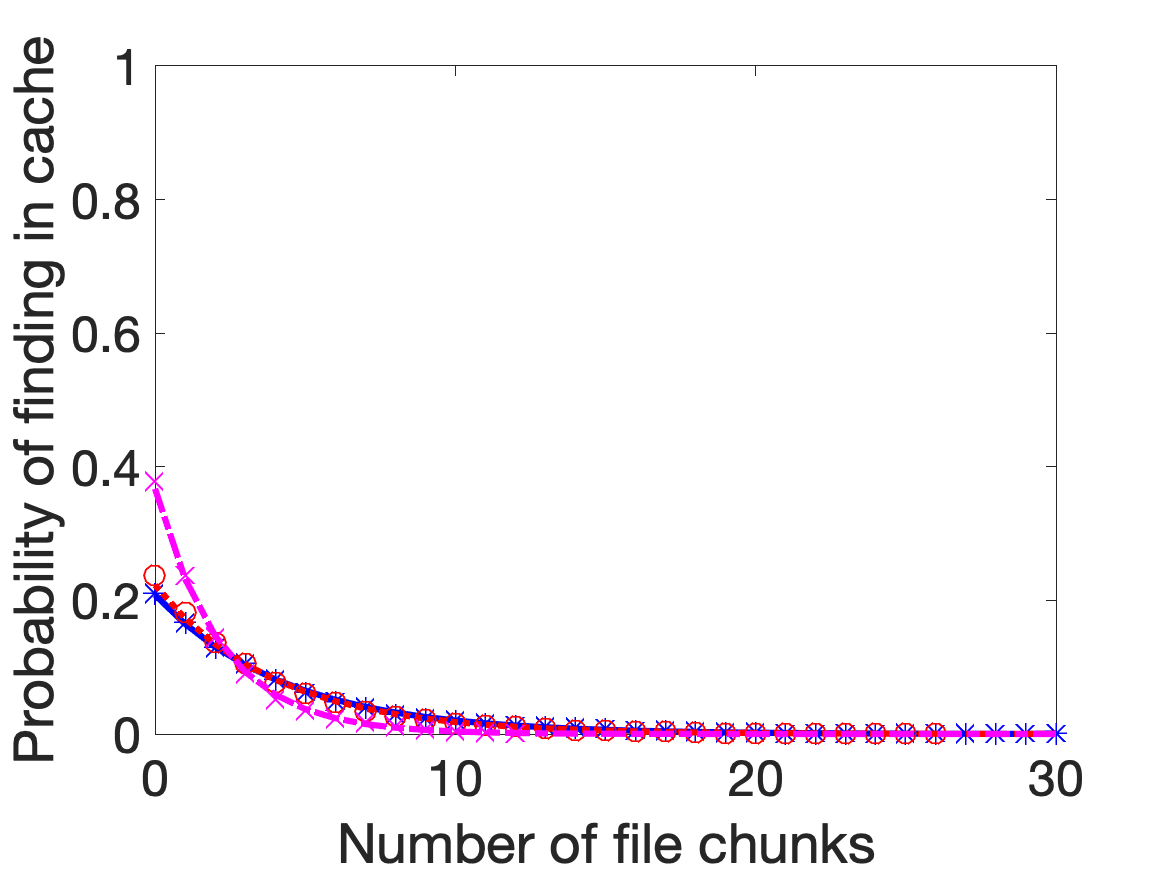}\caption{$\alpha=.8,\ C = 5000,\ c = 30$}\label{fig:cheest1}
\end{subfigure}
\begin{subfigure}[t]{0.33\textwidth}
\includegraphics[width = 1\textwidth]{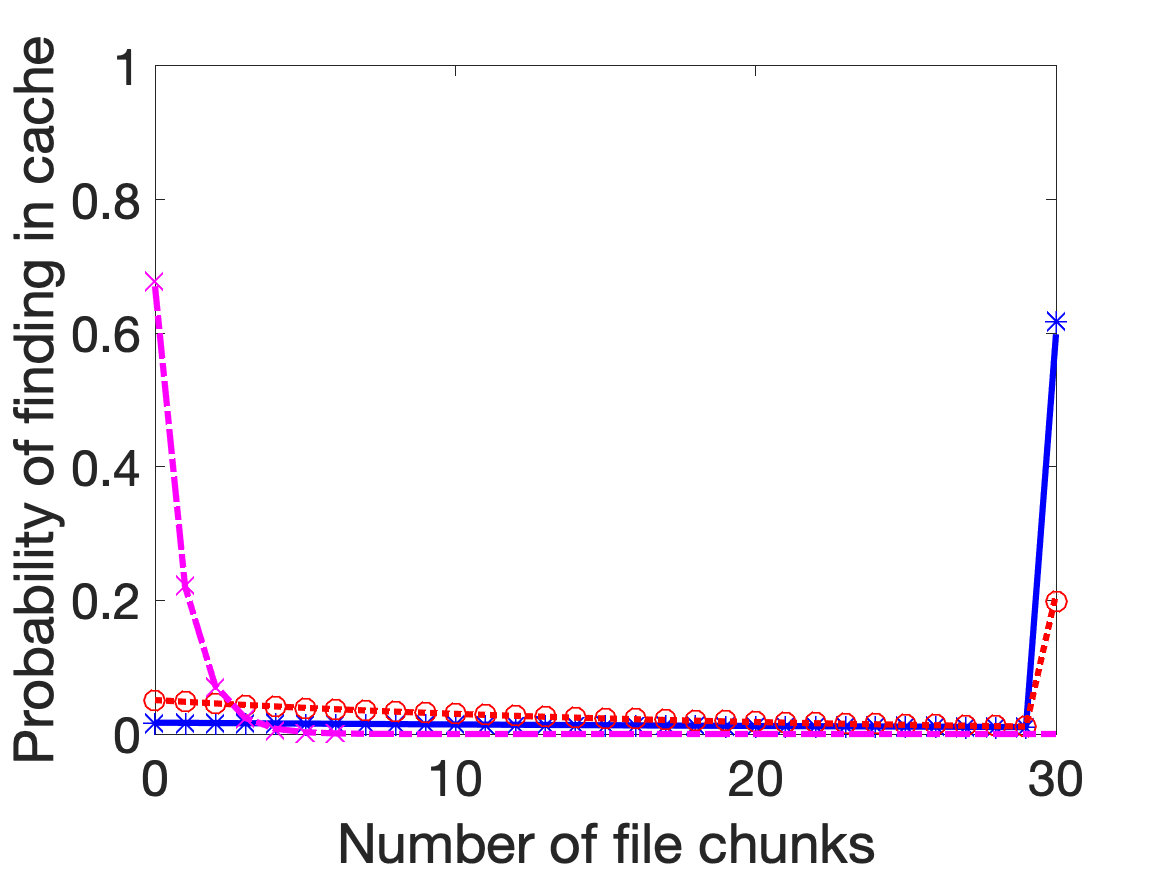}\caption{$\alpha=1.2,\ C = 5000,\ c = 30$}
\end{subfigure}
\begin{subfigure}[t]{0.325\textwidth}
\includegraphics[width = 1 \textwidth]{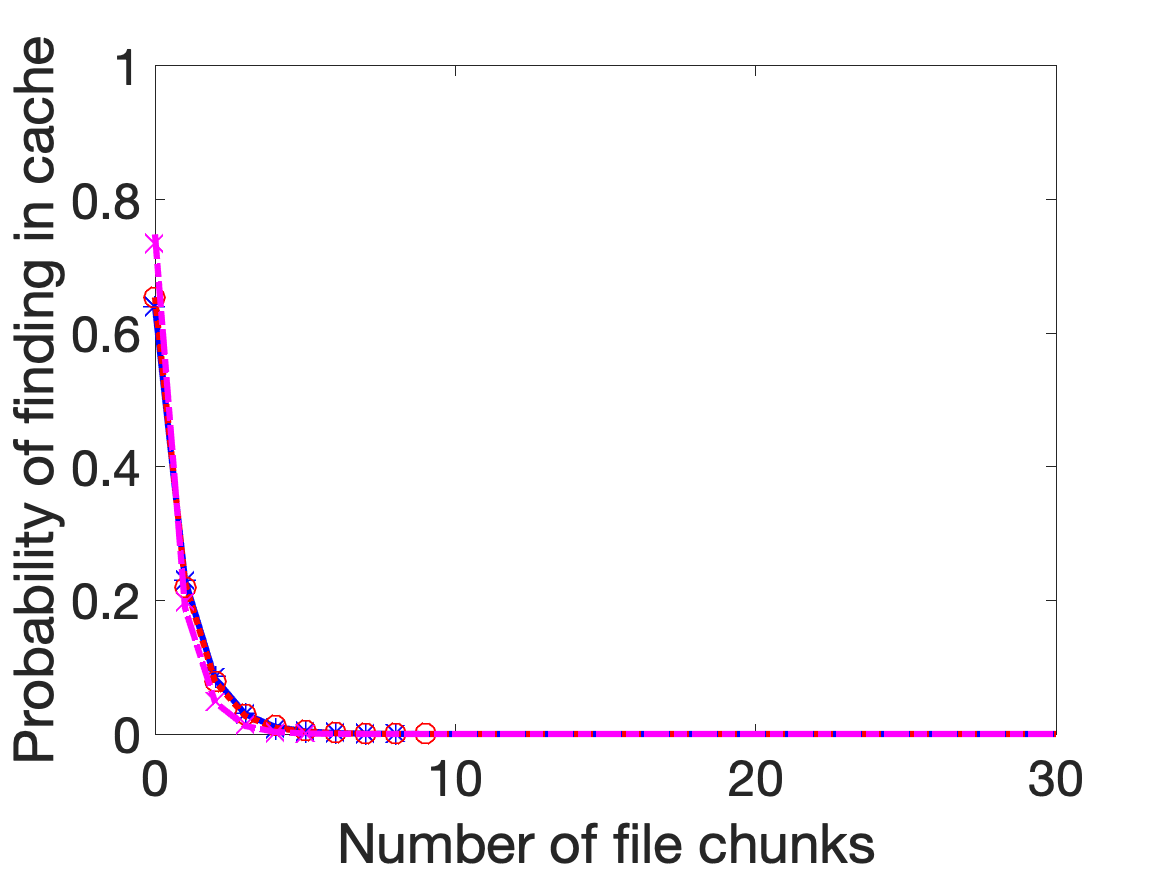}\caption{$\alpha=.8,\ C = 1000,\ c = 30$}
\end{subfigure}\\

\begin{subfigure}[t]{0.33\textwidth}
\includegraphics[width = 1\textwidth]{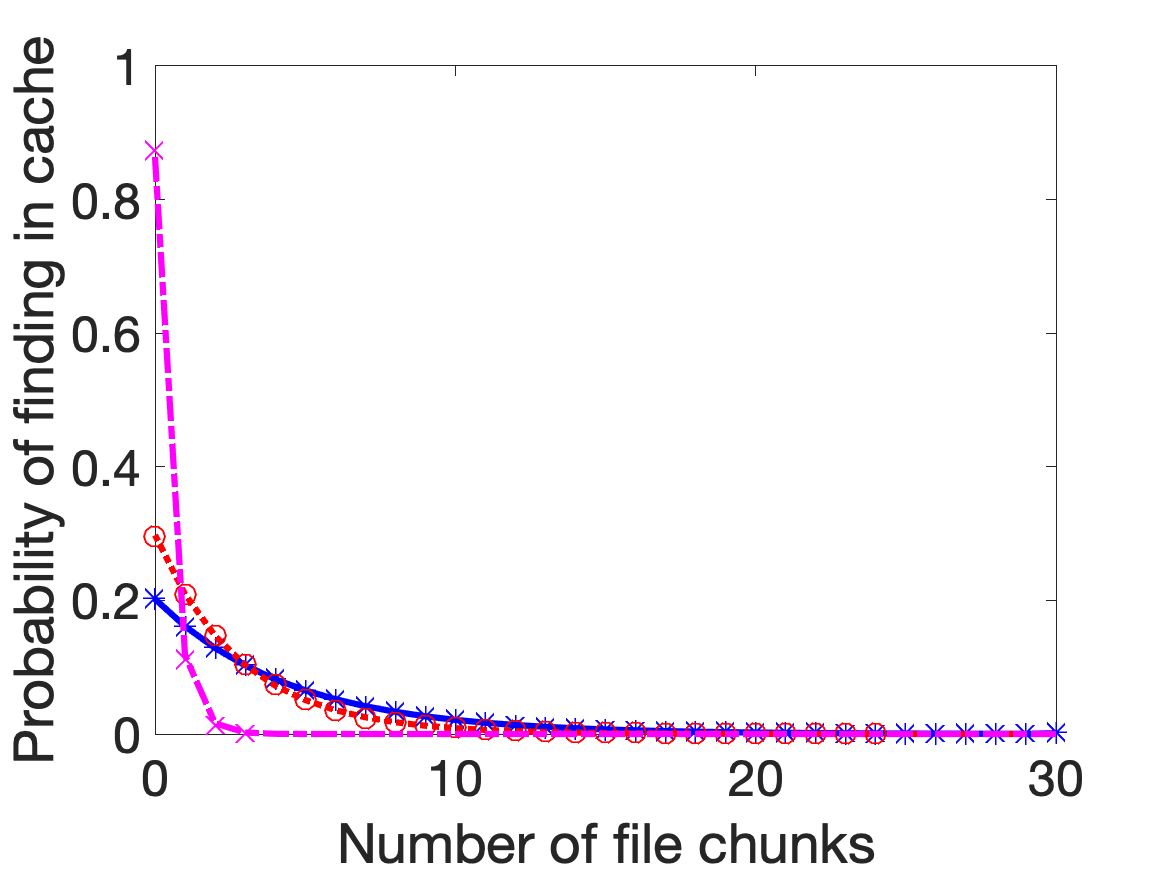}\caption{$\alpha=1.2,\ C = 1000,\ c = 30$}
\end{subfigure}
\begin{subfigure}[t]{0.33\textwidth}
\includegraphics[width = 1\textwidth]{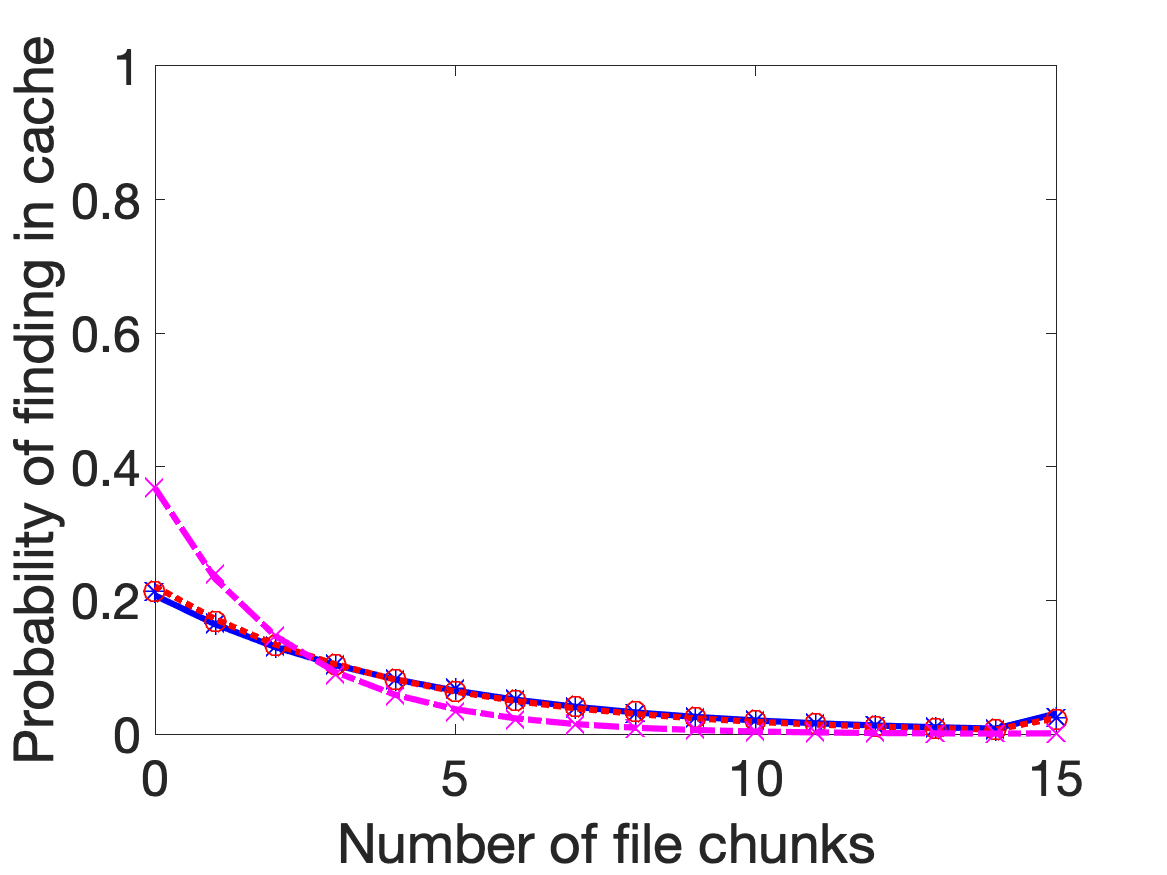}\caption{$\alpha=.8,\ C = 5000,\ c = 15$}
\end{subfigure}
\begin{subfigure}[t]{0.325\textwidth}
\includegraphics[width = 1\textwidth]{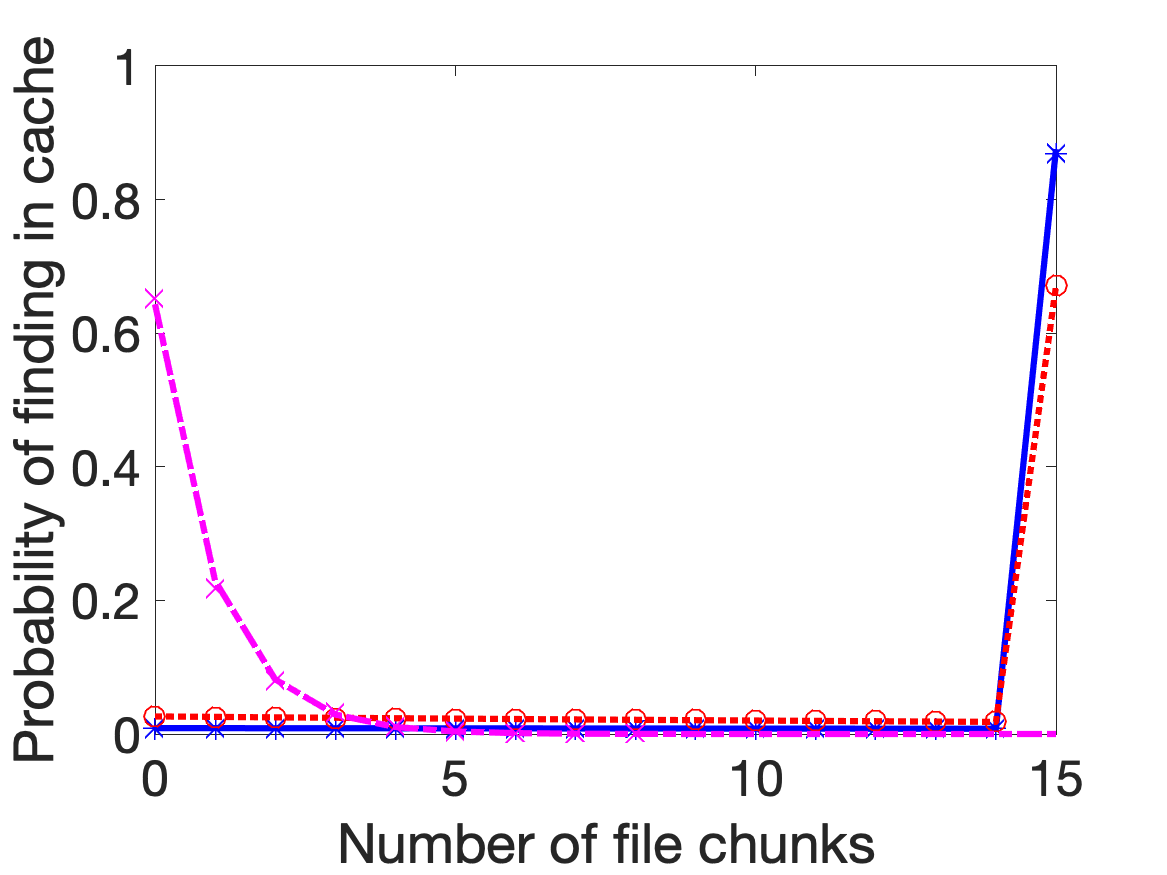}\caption{$\alpha=1.2,\ C = 5000,\ c = 15$}
\end{subfigure}\\

\begin{subfigure}[t]{0.33\textwidth}
\includegraphics[width = 1\textwidth]{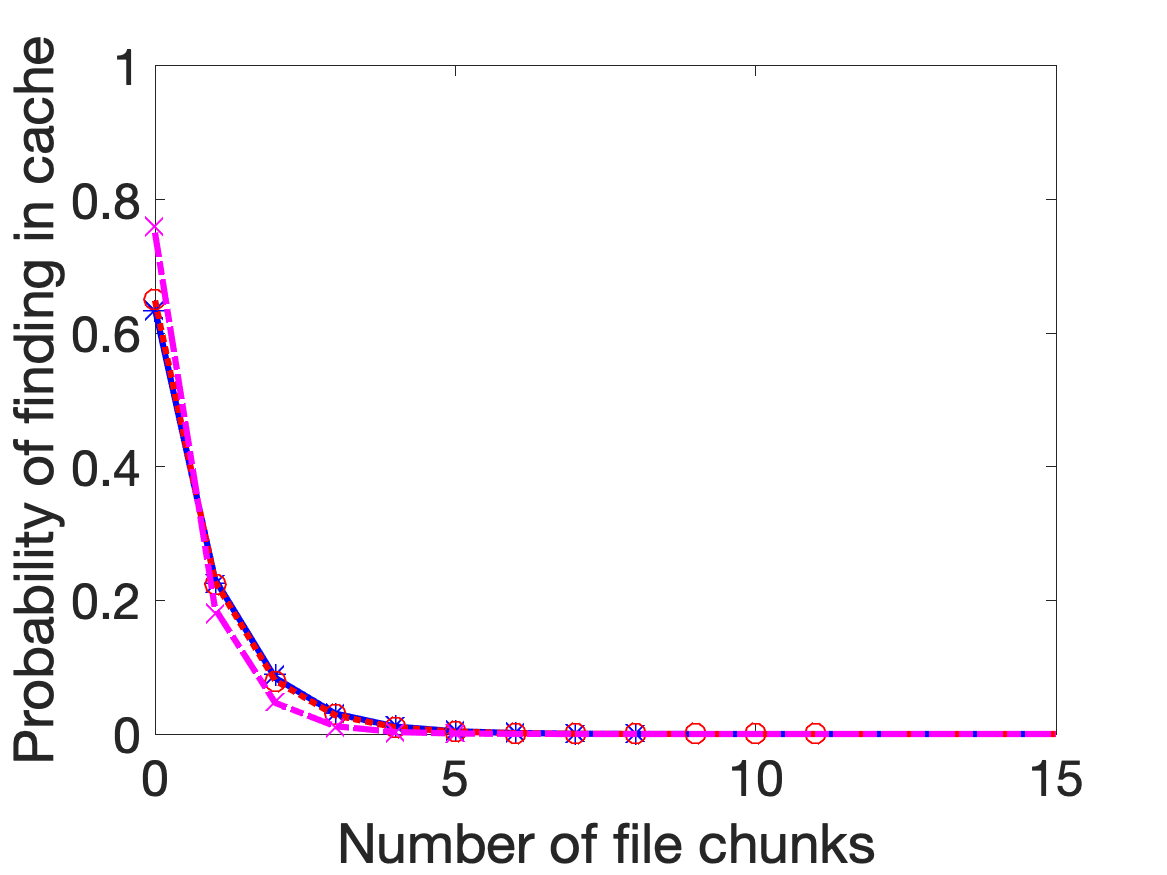}\caption{$\alpha=.8,\ C = 1000,\ c = 15$}
\end{subfigure}
\begin{subfigure}[t]{0.33\textwidth}
\includegraphics[width = 1\textwidth]{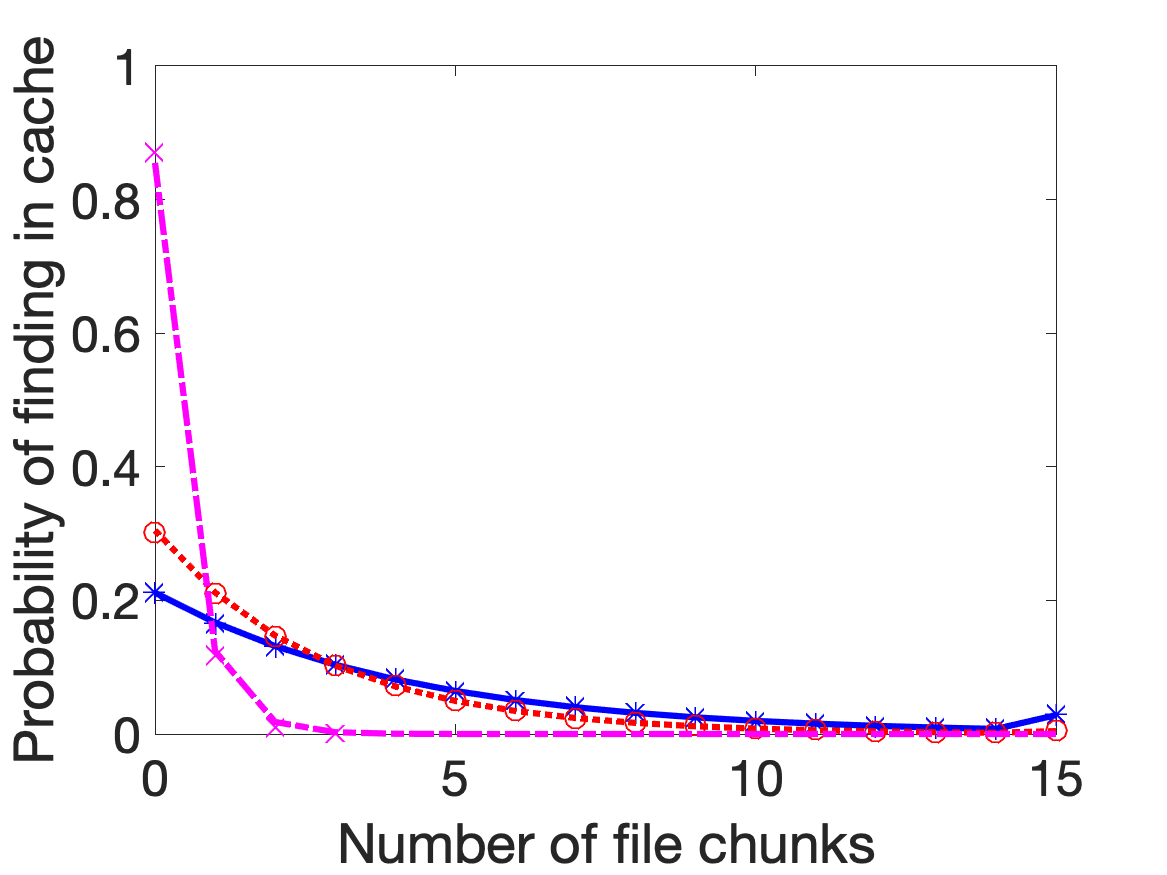}\caption{$\alpha=1.2,\ C = 1000,\ c = 15$}
\end{subfigure}
\begin{subfigure}[t]{0.325\textwidth}

\includegraphics[width = .85\textwidth, angle=-90, origin=c,trim={-4cm 0cm 0cm 0cm}, clip]{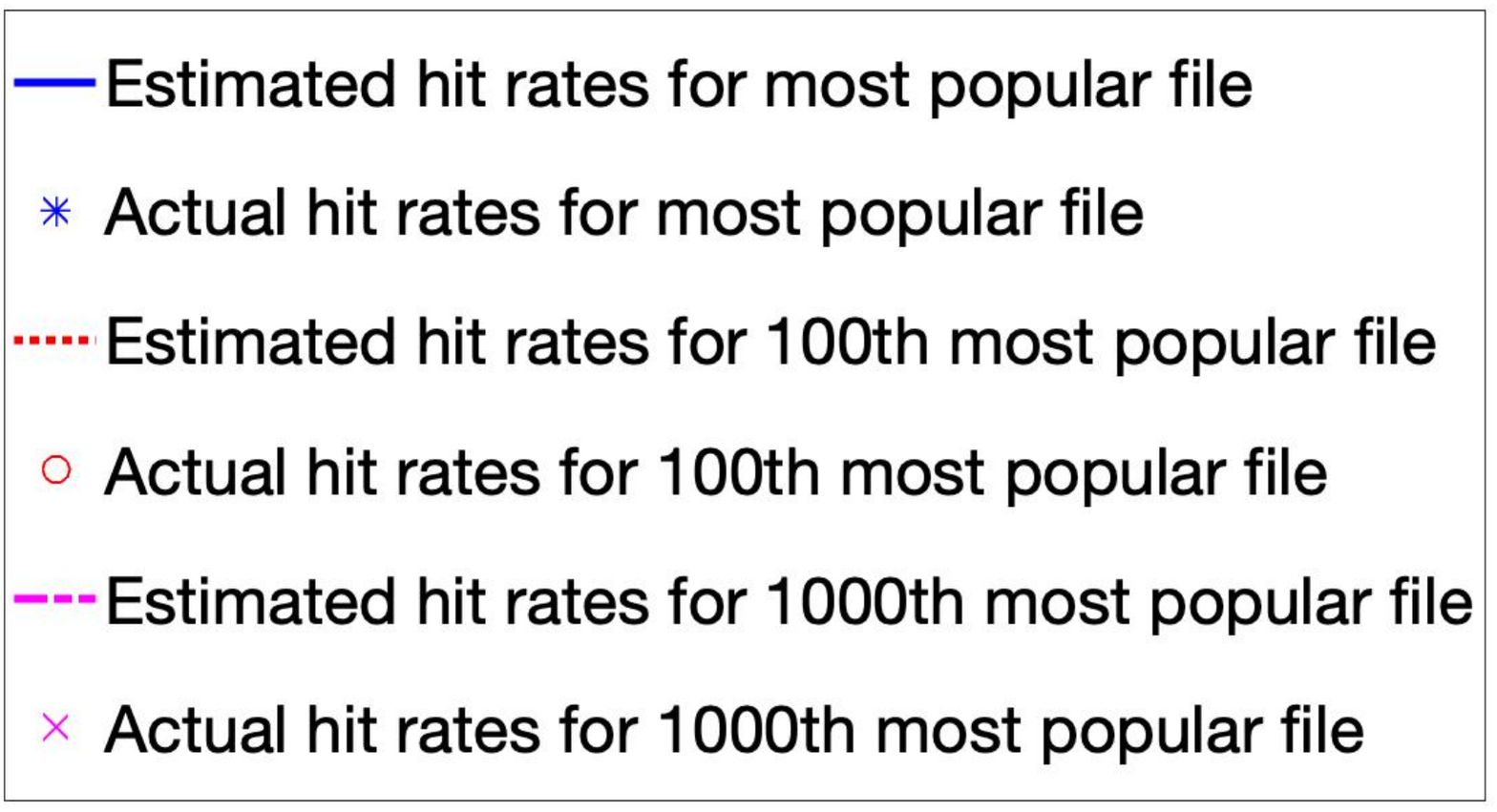}\caption{Legend}
\end{subfigure}
\caption{Comparison of true (markers) and estimated (lines) hit rates for a simulated caching system. The estimated hit rates are nearly perfect.}\label{fig:cheest}
\end{figure*}

The notion of hit rate in this context is slightly different than the hit rate in the characteristic time approximation \cite{fagin1977asymptotic, che2002hierarchical} in that we are no longer just concerned with the proportion of requests which find the requested file in the cache. 
Since the system can store partial files, we are interested in how many chunks of the requested file are found in the cache.
Consider an arbitrary file $i$. We are able to estimate the probability for finding exactly $k,\,0\leq k \leq s(i)$ chunks of file $i$ in the cache using Equation \eqref{jincache}. Figure \ref{fig:cheest} displays the estimated (solid lines) and simulated (markers) hit rates over a variety of parameters. Each point represents that probability (y-axis) of finding the number of chunks indicated on the x-axis. The blue, red, and pink represent the 1, 100, 1000, most popular files, respectively.
For example, if we look at the blue line at 5 on the x-axis in Figure \ref{fig:cheest1}, this represents the estimated probability of there being 5 cached chunks of the most popular file when it is requested and the star on top indicates the corresponding simulated value.
These figures show near perfect alignment between the estimated hit rates and true hit rates.
In particular, the largest difference between predicted and simulated values is .016 indicating that the approximations establish in Section \ref{sec:newche} are valid. 

\section{Numerical Evaluation of {g}LRU} \label{sec:numeval}
In this section, we compare gLRU with several other replacement policies. 
Section \ref{sec:videoStream} presents a comparison with LRU on a set of synthetic traces.
The results indicate the gLRU outperforms LRU on a variety of performance metrics in a video streaming setting.
In addition, we discuss some preliminary comparisons with two alternative policies, AdaptSize \cite{berger2017adaptsize} and a segment-based generalization of the LRU replacement policy which we will refer to as segLRU \cite{wu2001segment}.
See Section \ref{sec:AlternativePolicies}, for an outline of these two alternative policies. 
A brief discussion of preliminary comparisons on synthetic trace data is given at the end of Section \ref{sec:videoStream} and a more developed analysis is given comparing the results on production trace data in Section \ref{sec:trace}.
For the sake of reproducibility the code used to generate all figures and tables can be found at: \emph{https://github.com/EricFriedlander/gLRU}.

\subsection{VoD Latency for LRU and {gLRU} on Synthetic Trace Data}\label{sec:videoStream}
We now provide numerical simulations comparing the performance of the LRU and gLRU replacement policies.
All simulated systems consist of $N = 1,000$ video files.
The popularity of each video file is distributed according to a Zipf law with parameter $\alpha$ and the file sizes are distributed according to a censored Pareto distribution \cite{ramaswami2014modeling} with shape 2 and scale 300 (corresponding to an average length of 10 minutes) truncated so that no video is longer than 1 hour.
Each video file is broken down into chunks of length $L$ seconds.
When a video file is requested, the chunks are played sequentially, with each chunk consisting of $L$ seconds of video. 
In this section, we assume that the the popularity and file size distributions are independent.
However, the joint distribution is surely important and a topic of further work.
We give some  results in this direction in Appendix \ref{sec:dependence}.

Chunks retrieved from the cache are available to play immediately, while those not cached must be served by a single server FIFO queue as described in Section \ref{sec:netmodel}.
There is a start-up delay $d_s$ during which the video can be buffered before playing.
If the user attempts to view a file chunk which has not yet been served, they must wait for it to be processed and incur a delay.
The storage model is as described in Section \ref{sec:model}.
The  system is simulated under both the LRU and gLRU cache replacement policies. 

In our simulations, we study five performance metrics of interest,
\begin{enumerate}
\item[i)] $p_c$ - the proportion of file chunks retrieved from the cache. 
\item[ii)] $p_m$ - the proportion of requests in which no chunks are found in the cache.
\item[iii)] $T_w$ - the average amount of time required for each file to be retrieved, i.e., the download time of the file.
\item[iv)] $T_d$ - the average amount of time playback of a video is delayed. This is the re-buffering or the stall duration of the video and is a key metric in video streaming \cite{liu2015deriving}. This metric calculates the download time of each chunk and plays them in order with the play time of chunk $k$ being the maximum of the play time of chunk $k-1+L$ and the download time of chunk $k$. The difference of the play time of the last video chunk and $d_s + (s(i)-1)L$ gives the stall duration for file $i$.
\item[v)] $p_d$ - the proportion of requested videos which experience a nonzero stall duration.
\end{enumerate}

The input parameters are summarized in Table \ref{tab:paramDef}, which includes all definitions and values, and the system is simulated to convergence under the values described therein. 
Note that we do not explicitly give a processing rate $\mu$ because it is implicitly defined through the traffic parameter $\rho$.
\begin{table*}
\centering
\begin{tabular}{|c|c|c|}
\hline
Parameter & Definition & Values Used\\
\hline\hline
$\alpha$ & Zipf law parameter & 0.8, 1.2\\
\hline
$C_p$ & Cache sizes as proportion of total number of chunks & 0.1, 0.2\\
\hline 
$d_s$ & Delay before video playback. See Section \ref{sec:videoStream} & 3, 4 seconds\\
\hline
$L$ & The length of each video chunk & 1, 2, 3, 4 seconds\\
\hline
$\rho$ & \shortstack{The traffic intensity, i.e., the total arrival rate (weighted by file size)\\ divided by the processing rate (assuming no cache)} & 0.1, 0.5, 0.9\\
\hline
$r$ & Processing rate. Note that we assume each second of video corresponds to 3.13 MBps. & 1, 2, 10, 30 MBps\\
\hline
\end{tabular}\caption{List of parameters in LRU and gLRU simulations for VoD setting}\label{tab:paramDef}
\end{table*}
All combinations of parameters are simulated resulting in 384 separate trials. The reader should keep in mind that each configuration is only simulated once and thus the effects of the stochasticity in both the popularity distribution and file size distribution are not ``averaged out''. However, since there are 1000 files in each simulation, the effects should be minimal.

\begin{figure*}
\centering
\begin{subfigure}[t]{0.3\textwidth}
\includegraphics[width = 1\textwidth]{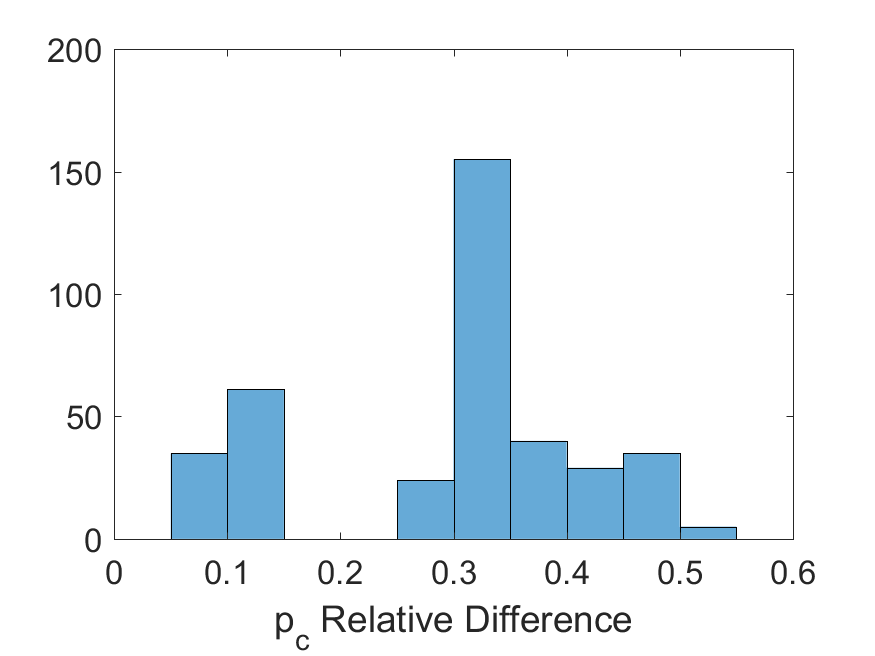}\caption{Proportion from Cache}
\end{subfigure}
\quad
\begin{subfigure}[t]{0.3\textwidth}
\includegraphics[width = 1\textwidth]{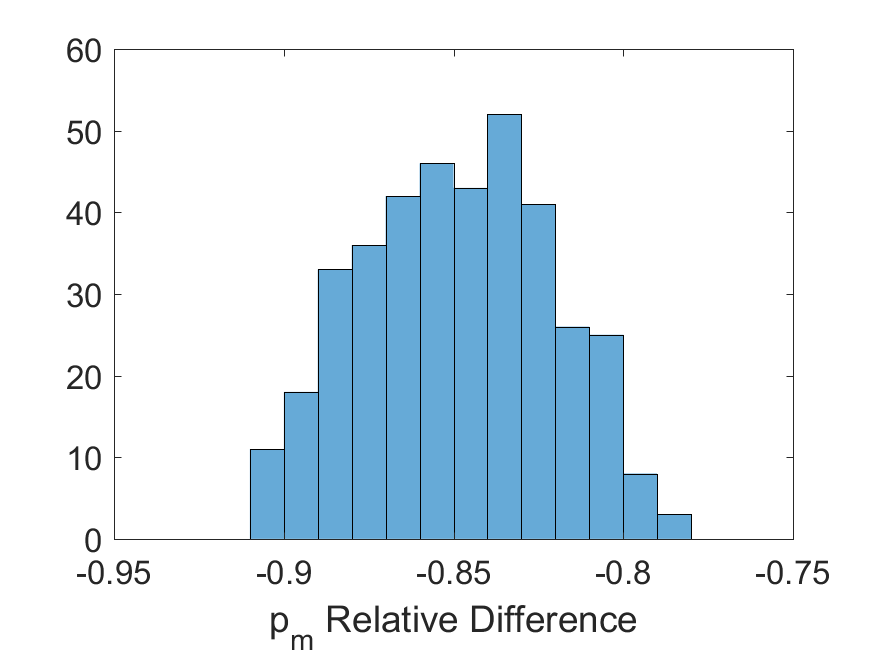}\caption{Cache Miss Rate}
\end{subfigure}\quad
\begin{subfigure}[t]{0.3\textwidth}
\includegraphics[width = 1 \textwidth]{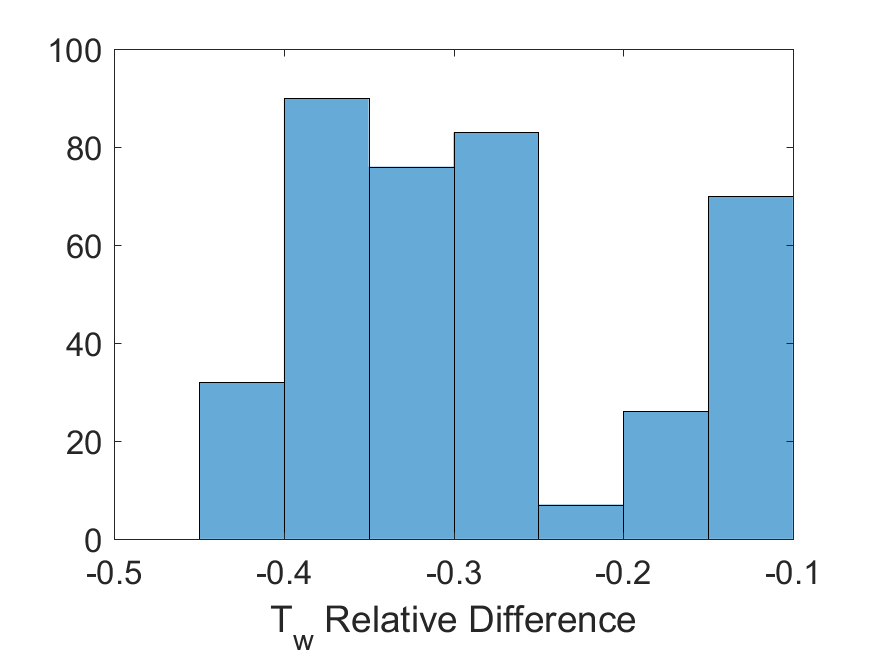}\caption{Download Time}
\end{subfigure}\\

\begin{subfigure}[t]{0.3\textwidth}
\includegraphics[width = 1\textwidth]{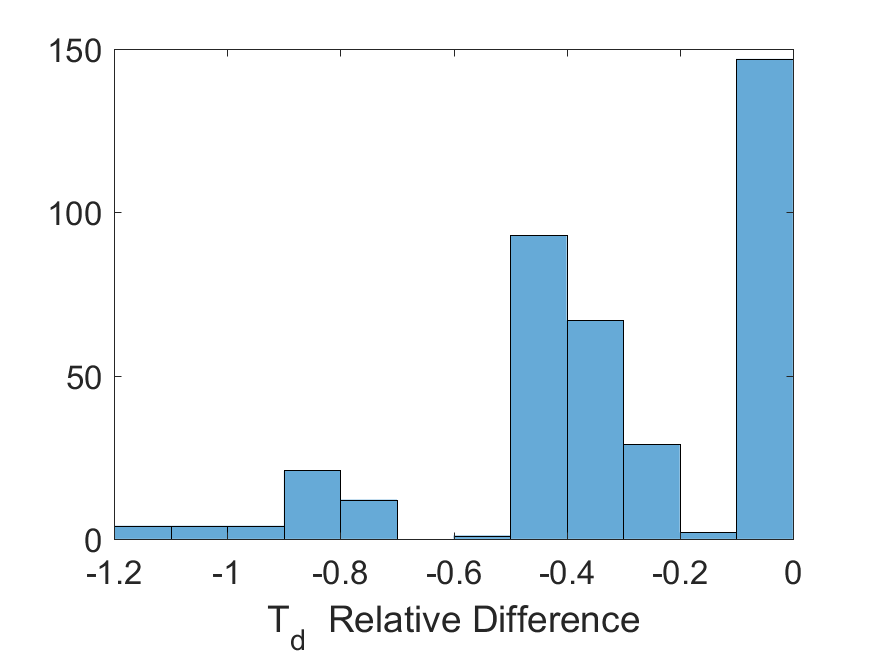}\caption{Delay Time}\label{fig:histtd}
\end{subfigure}\quad
\begin{subfigure}[t]{0.3\textwidth}
\includegraphics[width = 1\textwidth]{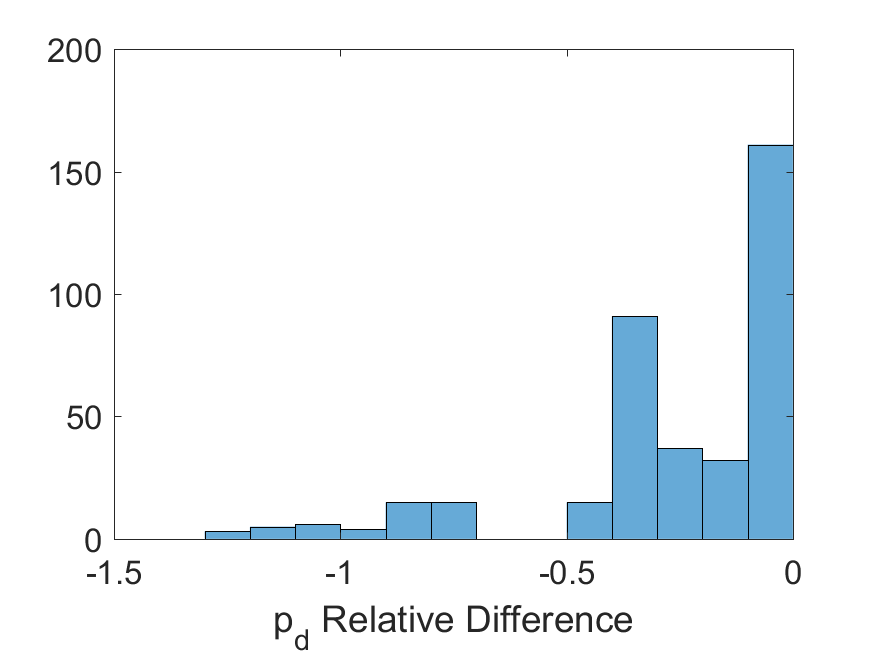}\caption{Proportion Delayed}\label{fig:histpd}
\end{subfigure}\caption{Histograms of relative performance improvement of gLRU over LRU. gLRU outperforms LRU in almost every instance.}
\label{fig:histimprovement}
\end{figure*}

In Figure \ref{fig:histimprovement}, we present the results of the simulations.
A histogram of the relative difference between gLRU and LRU for all performance metrics is presented. Specifically, the $x$-axes correspond to the difference between the specified performance metric under gLRU and LRU divided by the metric for LRU.
Histograms of the gross differences can be found in Appendix \ref{sec:AppFig}.

In Table \ref{tab:metricsum}, we present the results of comparing gLRU with LRU.
The columns ``Worse'' and ``Better'' represent the number of trials in which gLRU performs worse and better and ``Worst'', ``Best'', and ``Median'' give the worst, best, and median improvement of gLRU over LRU.
In the tables, positive numbers correspond to gLRU performing better.
The results are given in terms of relative improvement where the magnitude is computed as (gLRU-LRU)/LRU.
When considering $p_c$, $p_m$, and $T_w$,  gLRU outperforms LRU in every instance. For the delay focused metrics, $T_d$ and $p_d$, gLRU outperforms LRU in 237 and 240 out of 384 configuration, respectively. In the remaining cases there was no discernible difference in performance. 
\begin{table}
\centering
\begin{tabular}{|c|c|c|c|c|c|}
\hline
\shortstack{Performance\\Metric} & Worse & Better & Worst  &  Best & Median \\
\hline\hline
$p_c$ & 0	   &  384      &  7.3\% & 51.9\% & 31.9\% \\
\hline
$p_m$  &  0	   &  384      &  78.4\% & 91.0\% & 84.9\%\\
\hline
$T_w$ &  0	   &  384     &  11.0\% &  44.0\% & 30.8\%\\
\hline
$T_d$ & 0  &  237      &  0\% &  118.7\% & 32.0\%\\
\hline
$p_d$ & 0   &  240      &  0\% & 128.4\%  & 19.7\%\\
\hline
\end{tabular}
\caption{Comparison of gLRU to LRU shows that gLRU outperforms LRU in almost every trial. Rows with less that 384 trials indicate that there were ties.}
\label{tab:metricsum}
\end{table}

These results  indicate that the gLRU replacement policy is almost always superior to the LRU policy and results in shorter download times (30.8\% improvement), lower stall duration (32.0\% improvement), and more videos with non-zero stalls (19.7\% improvement). Since videos are watched sequentially (i.e., a user begins at the beginning and proceeds through the file pieces one by one), it make sense that gLRU would result in an improvement in the VoD case since this viewing of earlier video chunks provides time for chunks appearing later in the video to load.
Moreover, our results show that by employing a gLRU policy, a system designer is able to improve the user experience by increasing the number of files which are partially stored in the cache even is non-VoD settings due to the improvement in the download time.
The ultimate result is a system with fewer and shorter delays and shorter download times.

In addition, we compared gLRU with segLRU and AdaptSize (cf. Section \ref{sec:AlternativePolicies}) on synthetic traces with the same parameters as discussed above, although with smaller numbers of files and fewer numbers of requests.
These preliminary results show that gLRU outperforms AdaptSize most of the time, while significantly under-performing segLRU.
It is not clear exactly what is causing the difference in performance, as the production trace evaluation in Section \ref{sec:trace} show a significant performance improvement of gLRU over segLRU.
Furthermore, we are not troubled by these results as we believe that gLRU will prove useful when incorporated, instead of LRU, in more complicated replacement policies and content delivery architectures.

\subsection{Alternative Policies}\label{sec:AlternativePolicies}
\paragraph{AdaptSize} AdaptSize is a probabilistic admission policy in which a file is admitted into the cache with probability $e^{-\text{object size}/c}$.
Larger objects are admitted with lower probability and the parameter $c$ is tuned to maximize the object hit rate (OHR), defined as the probability that a requested file is found in the cache.
In particular, given a $c$ and estimates on the arrival rate for the requests for each file, one can estimate the probability that a given file will be found in the cache. 
One can then use these probabilities to compute the OHR as a function of $c$ and then optimize.
This process is performed in windows, recomputing $c$ after a given number of file requests.
We refer the reader to \cite{berger2017adaptsize} for a more in-depth description.
In all simulations, we take our window size to be 1,000 requests.

\paragraph{segLRU} This segment-based caching policy relies on breaking up chunked files into segments with segment $i$ containing $2^{i-1}$ chunks.
It follows that chunks which occur early in a video are contained in very small segments while those which occur later are contained in large segments.
The cache is then broken into two stacks.
The first stack is devoted to the first $K_{min}$ segments of each file and the second devoted to the later segments.
The first stack operates under a simple LRU replacement policy while the second uses a thresholding approach.
In particular, each segment receives a score dependent on the time of its last request and position in the video file and is added to the cache only if its score is higher than another segment currently in the cache.
In this case, the lowest ranked segment is evicted to make space for the one with the higher score.
For more information we refer the reader to \cite{wu2001segment}.
Throughout this section we take each stack as half the size of the cache in the other policies and set $K_{min}$ so that one quarter of all chunks are directed to the first cache.

\subsection{VoD Latency for LRU, segLRU, and AdaptSize on Production Trace Data}\label{sec:trace}
We now compare gLRU with segLRU and AdaptSize on MSR Cambridge Traces obtained from the SNIA IOTTA repository.
There are 36 I/O traces from 36 different volumes on 13 servers.  
For more information see \cite{narayanan2008write}.
The data includes the timestamp of each request, the file size of each request, and whether the entry is a read or write request.
For the comparisons below we only consider read requests.
No file identifiers are given, so we take each file size as the unique identifier.
We test our policies on two of the 36 traces.
Specifically, we select the first media trace and the first web trace which include 143,973 and 606,487 read requests, respectively.
These were selected both because they were large but also because of the distribution of file requests.
Namely, in the second media trace, over 95\% of the files requests were for files of the same size and recall that we take the file size as the unique identified.
In contrast, the first web trace had more than half of the files account for at least 1\% of the traffic each.
We assume the same system as in Section \ref{sec:videoStream} in which files are retrieved immediately from the cache if available, otherwise they are served from a FIFO queue.
Several combinations of parameters are considered.
We consider settings in which the chunk size is chosen so that the average file has 10, 50, and 100 chunks.
Furthermore, we consider a range of eight cache sizes capable of storing between 1\% and 50\% of the total files in the system.
This range of parameters results in 48 separate simulations per replacement policy.
The processing rate and start-up delay are chosen so that the traffic rate on the queue is manageable.
In order to initialize the queue and cache, we run our simulation through each trace once and use the end state of the queue and cache as the initial condition for all future simulations.

Tables \ref{tab:gvseg} and \ref{tab:gvas} present the results of comparing gLRU with segLRU and AdaptSize, respectively.
The columns ``Worse'' and ``Better'' represent the number of trials in which gLRU performs worse and better and ``Worst'', ``Best'', and ``Median'' give the worst, best, and median improvement of gLRU over the competing replacement policy.
In the tables, positive numbers correspond to gLRU performing better.
The results show that gLRU outperforms segLRU in almost every trial.
In addition, gLRU performs favorably to AdaptSize, having lower delay time in 45 out of 48 trials, and a lower proportion in delayed requests in 36 out of 48 trials.
Furthermore, the gains in performance for the cases in which gLRU performs better seem to be quite large while they are much smaller in cases where AdaptSize outperforms gLRU.
\begin{table}
\centering
\begin{tabular}{|c|c|c|c|c|c|}
\hline
\shortstack{Performance\\Metric} & Worse & Better & Worst  &  Best & Median \\
\hline\hline
$p_c$ & 3	   &  45      &  -.71\%& 231\% & 1.79\% \\
\hline
$p_m$  &  3	   &  43      &  -119\% &  89.6\% & 53.7\%\\
\hline
$T_w$ &  3	   &  45      &  -140\% &  97.6\% & 72.8\%\\
\hline
$T_d$ & 2  &  46      &  0\% &  99.8\% & 78.2\%\\
\hline
$p_d$ & 3   &  45      &  17.5\% & 97.9\%  & 52.2\%\\
\hline
\end{tabular}
\caption{Comparison of segLRU to gLRU shows that gLRU outperforms segLRU in almost every trial. Rows with less that 48 trials indicate that there were ties.}
\label{tab:gvseg}
\end{table}

\begin{table}
\centering
\begin{tabular}{|c|c|c|c|c|c|}
\hline
\shortstack{Performance\\Metric} & Worse & Better & Worst  &  Best & Median \\
\hline\hline
$p_c$ & 21  &  27      &  -1.49\% & 41.9\% & .16\% \\
\hline
$p_m$  &  0	   &  48      &  23.3\% &  95.1\% & 68.4\%\\
\hline
$T_w$ &  0	   &  48      &  72.4\% &  99.4\% & 96.1\%\\
\hline
$T_d$ & 3   &  45      & -28.3\% & 95.1\% & 35.0\%  \\
\hline
$p_d$ & 12  &  36      &  -35.3\% & 97.3\%  & 9.57\%\\
\hline
\end{tabular}
\caption{Comparison of AdaptSize to gLRU shows that gLRU performs favorably compared to AdaptSize.}
\label{tab:gvas}
\end{table}

%% file: conclusion.tex
\section{Conclusions and Future Work}\label{sec:conclusion}
In this work, we present a generalization for the LRU cache replacement policy, called gLRU.
We first establish a generalization of  the characteristic time approximation to this new policy which can be used to accurately approximate the distribution of the number of chunks of a file in the cache in the steady state.
The results of Section \ref{sec:chevalid} show that the approximation is, indeed, quite accurate.
We then provide numerical results demonstrating that gLRU outperforms gLRU on a variety of performance metrics including download time and stall duration in a VoD setting.
Our results indicate that gLRU  outperforms LRU in  non-VoD settings as well because of its superior performance on download time.
Furthermore, preliminary comparisons with alternative replacement policies indicate that gLRU will be a valuable tool in improving the efficiency of content delivery architectures.

Exploring generalizations of gLRU, such as how segment-based additions can  be exploited to hone the reactivity of the algorithm present an interesting direction for future research.
Furthermore, it remains to be seen how more complicated architectures can be improved from the addition of gLRU in place of LRU caches.
It is also worth thinking about how this policy can fit into an adaptive streaming framework  \cite{sanchez2011idash,muller2011test,8393453,elgabli2019fastscan}, where each chunk could be fetched at different quality levels. Exploring a trade-off between the average quality at which adaptive video is streamed and the stall duration with an online caching algorithm is an important direction for the future.

%% file: apdx.tex
\appendices

\section{Hueristic Analysis  for gLRU($d$)}\label{sec:newched}

In this subsection, we extend the analysis of Section \ref{sec:newche} to gLRU($d$). 
Take  $i$, $X_n(t)$, $s(i)$, $t_C$, $d(i,t_C)$, and $\tau_{n,k}$ as in Section \ref{sec:newche}.
For a given $j\in\NN$, define $k_{req}(j)$ such that $H_{k_{req}(j)-1,-d}< j\leq H_{k_{req}(j),-d}$ where $H_{n,p} = \sum_{k=1}^n k^{-p}$ are generalized Harmonic numbers \cite{choi2011some}.
The probability of finding at least $j$ chunks of file $n$ in the cache is equal to the probability that the last $k_{req}(j)$ inter-arrival times, looking back from the current time, for file $n$ are less than $t_C$.
This can be approximated as follows,
\begin{align}\label{eqn:hitprobapdx}
h_j(n,t_C) \doteq \PP(\tau_{n,1}<t_C,\ldots, \tau_{n,j_{req}}<t_C) = (1-e^{q(n)t_C})^{k_{req}(j)}.
\end{align}

Let $Y_i,i\in\{1,\ldots,N\}$   be as in Section \ref{sec:newche}, the number of cached chunks of file $i$ when the system is in steady state.
Recalling that $d(i, t_C)$ denotes the expected number of file $i$ chunks in the cache when the system is in steady state, it then follows that
\begin{align*}
d(i,t_C)
= \sum_{j=1}^{s(i)}\PP(Y_i\geq j)
= \sum_{j=1}^{s(i)}(1-e^{q(i)t_C})^{k_{req}(j)}.
\end{align*}
The proposed approximation of $X_n(t)$ is then given as
\begin{align*}
\ti X_n(t) = \sum_{i = 1,i\neq n}^Nd(i,t).
\end{align*}
As in the derivation of \eqref{eqn:origche2}, it is sufficient to consider
\begin{align*}
\ti X(t) = \sum_{i = 1}^Nd(i,t)
\end{align*}
and one can then estimate $t_C$ by setting the expected value of $\ti X$ to $C$ and solving for $t_C$. 
This amounts to solving the following equation,
\begin{align*}
C = \E\ti X = \sum_{i=1}^N\sum_{j=1}^{s(i)}(1-e^{q(i)t_C})^{k_{req}(j)}.
\end{align*}
As in Section \ref{sec:newche}, and using $h$ as defined in \eqref{eqn:hitprobapdx} this can then be used to find other metrics of interest.

\section{VoD Gross Difference}\label{sec:AppFig}
\begin{figure*}[ht]
\centering
\begin{subfigure}[t]{0.3\textwidth}
\includegraphics[width = 1\textwidth]{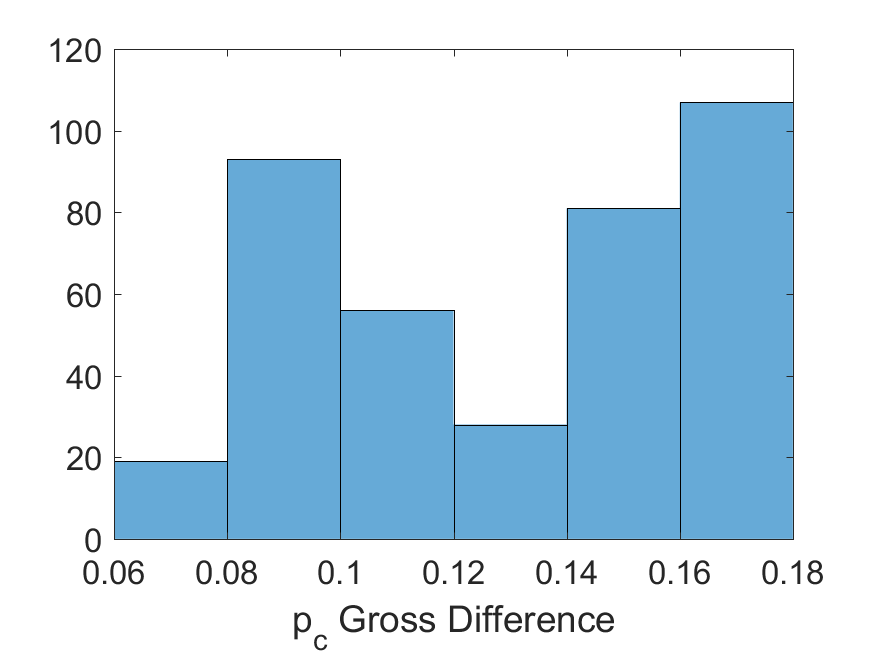}\caption{Proportion from Cache}
\end{subfigure}
\quad
\begin{subfigure}[t]{0.3\textwidth}
\includegraphics[width = 1\textwidth]{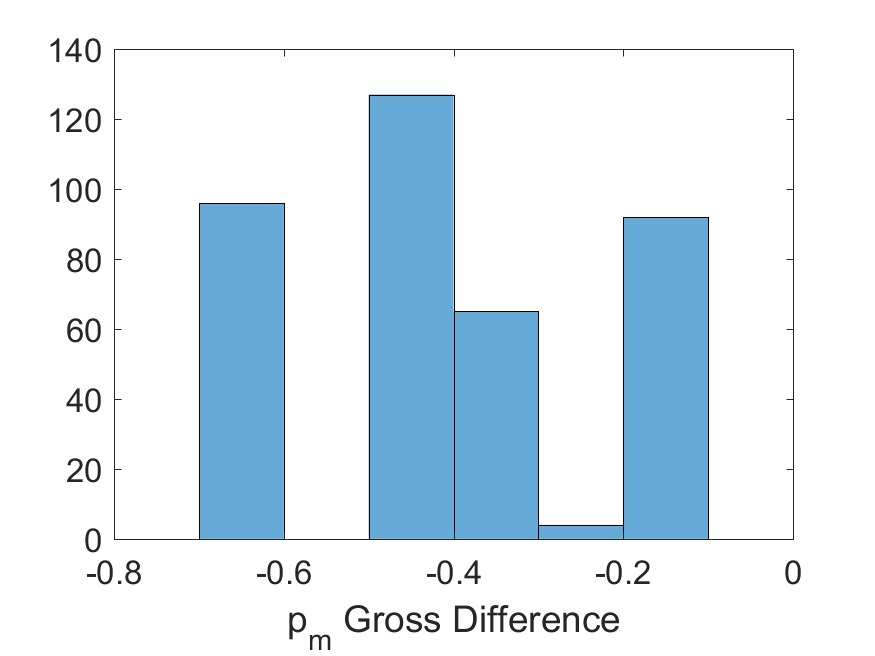}\caption{Cache Miss Rate}
\end{subfigure}\quad
\begin{subfigure}[t]{0.3\textwidth}
\includegraphics[width = 1 \textwidth]{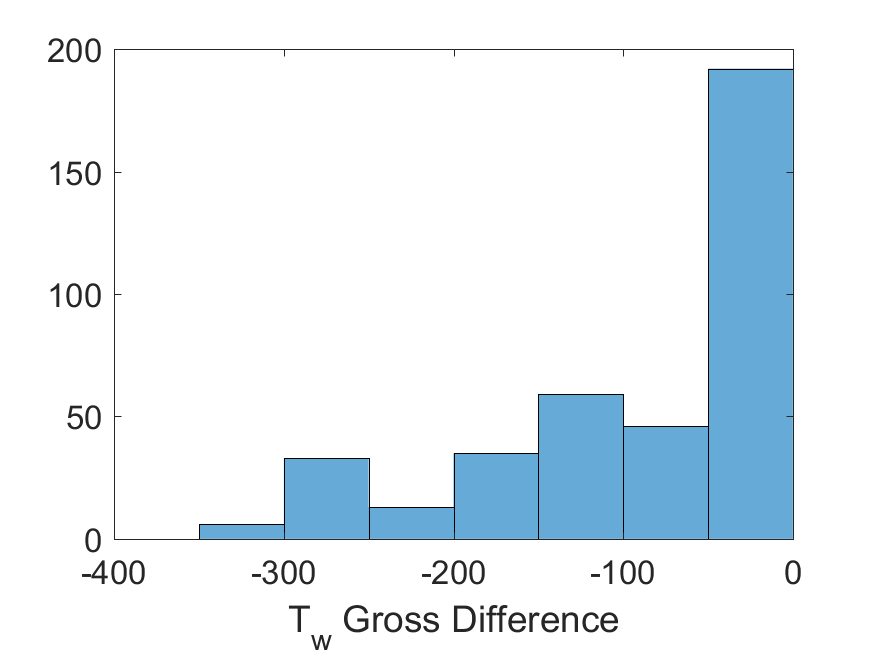}\caption{Download Time}
\end{subfigure}\\

\begin{subfigure}[t]{0.3\textwidth}
\includegraphics[width = 1\textwidth]{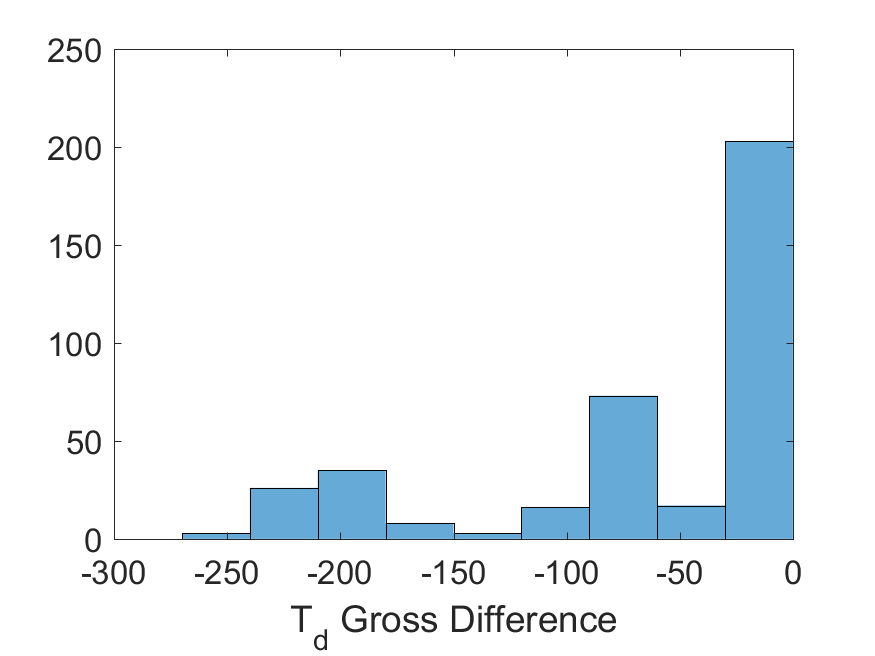}\caption{Delay Time}\label{fig:histtdgr}
\end{subfigure}\quad
\begin{subfigure}[t]{0.3\textwidth}
\includegraphics[width = 1\textwidth]{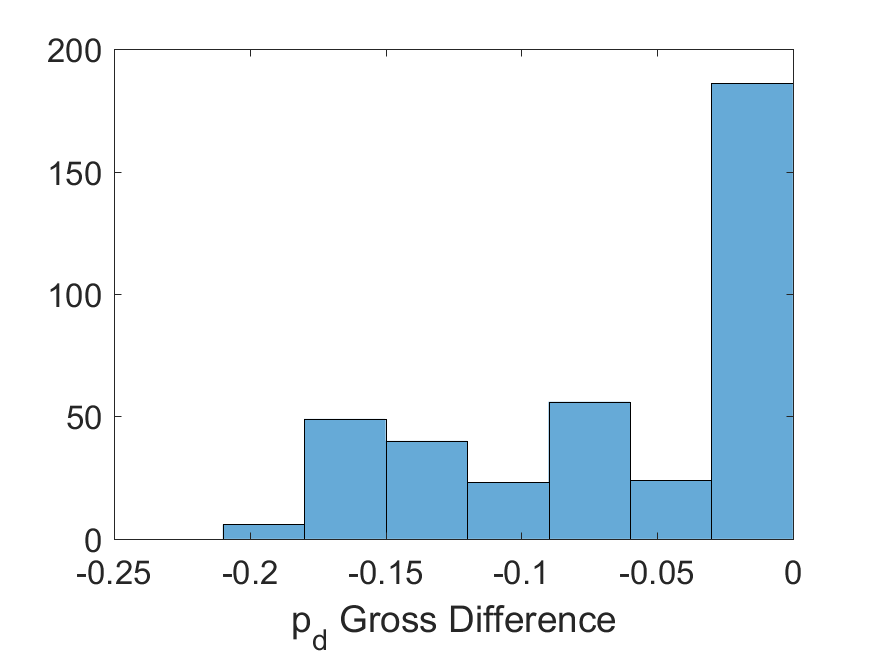}\caption{Proportion Delayed}\label{fig:histpdgr}
\end{subfigure}\caption{Histograms of gross performance improvement of gLRU over LRU. gLRU outperforms LRU is almost every instance. Even in \ref{fig:histtdgr} and \ref{fig:histpdgr} the difference in performance when LRU outperforms gLRU are nearly negligible.}
\label{fig:histimprovementgr}
\end{figure*}

In order to supplement the results of Section \ref{sec:videoStream}, we also present the gross improvement of gLRU over LRU.
Figure \ref{fig:histimprovementgr} presents histograms of the gross improvement of gLRU over LRU for each performance metric of interest.
In each histogram the $x$-axes corresponds to the difference between the specified performance metric under gLRU and LRU.
The figures back-up the results of Section \ref{sec:videoStream} in demonstrating the improved performance of gLRU over LRU for all performance metrics considered.

\section{Joint Popularity and File Size Distribution}\label{sec:dependence}
We now briefly consider the impact of the joint distribution of the popularity and file size on each of our performance metrics.
Modeling this joint distribution can be quite complicated.
In particular, specifying a probability measure such that the marginal distributions are Zipf for popularity and Pareto for file size while imposing a given correlation structure between the two is a nontrivial problem on its own.

In this section, we consider of the effect of strong positive and negative correlation between popularity and file size through simulation.
The system, as described in Section \ref{sec:videoStream}, is simulated with one major change.
In order to induce strong positive correlation we simply assign the highest popularities to the largest files and vice versa to induce negative correlation.
The system was then simulated for all of the combinations of parameters described in Table \ref{tab:paramDef} for both cases resulting in 384 cases for each.

Table \ref{tab:posvneg} presents a comparison of the positive and negative correlation cases for both the gLRU and LRU settings.
For each of the performance metrics described in Section \ref{sec:videoStream}, we present the average relative difference in performance positive vs. negative correlation (i.e. (positive metric - negative metric)/positive metric).
\begin{table}
\centering
\begin{tabular}{|c|c|c|}
\hline
Metric & gLRU & LRU \\
\hline \hline
$p_c$ & -.41     & -.47       \\
\hline
$p_m$ & 4.03	 & 2.16     \\
\hline
$T_w$ & 9.52     & 8.02 \\
\hline
$T_d$ & 4.99     & 4.28 \\
\hline
$p_d$ & 2.42         & 1.37     \\
\hline
\end{tabular}
\caption{Mean relative difference in performance for the positive vs. negative correlation setting. Values are computed as (positive correlation - negative correlation)/negative correlation for each metric of interest.}
\label{tab:posvneg}
\end{table}
The results show that the case of negative correlation vastly outperforms the case of positive correlation in both when employing either a gLRU or LRU replacement policy.
%
This is because in the case of negative correlation, most popular files are smaller and thus more of them can be stored in the cache improving all the metrics as compared to when there is positive correlation. 


In addition, we note that gLRU outperforms LRU in both the positive and negative correlation case.
In Table \ref{tab:posneg}, we present the mean relative performance improvement for gLRU vs LRU in all performance metrics of interest in both the correlation settings.
\begin{table}
\centering
\begin{tabular}{|c|c|c|}
\hline
Metric & Positive Correlation & Negative Correlation \\
\hline \hline
$p_c$ & .42     & .21        \\
\hline
$p_m$ & -.57 	&  -1.51    \\
\hline
$T_w$ & -.10    & -.82\\
\hline
$T_d$ & -.14    & -.69 \\
\hline
$p_d$ & -.16    & -.46     \\
\hline
\end{tabular}
\caption{Mean relative performance improvement of gLRU over LRU in the case of both positive and negative correlation for each of the performance metrics outlined in Section \ref{sec:videoStream}}
\label{tab:posneg}
\end{table}
While the results are much more pronounced in the case of negative correlation, it is clear that gLRU outperforms LRU in both correlation cases.
For a more granular look at the data see Appendix \ref{app:corrhist}.

\section{Correlation Histograms}\label{app:corrhist}
\begin{figure*}[ht]
\centering
\begin{subfigure}[t]{0.3\textwidth}
\includegraphics[width = 1\textwidth]{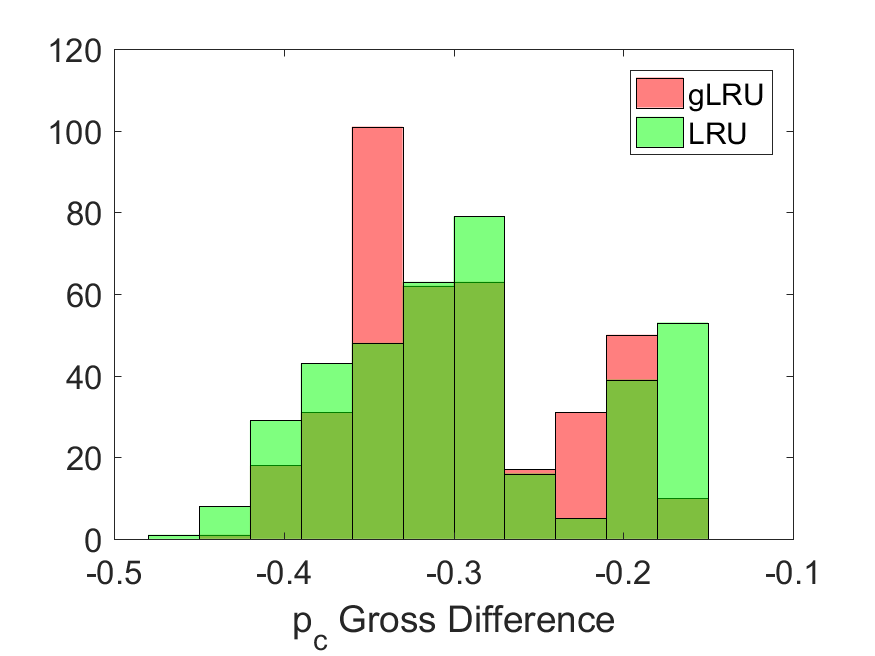}\caption{Proportion from Cache}
\end{subfigure}
\quad
\begin{subfigure}[t]{0.3\textwidth}
\includegraphics[width = 1\textwidth]{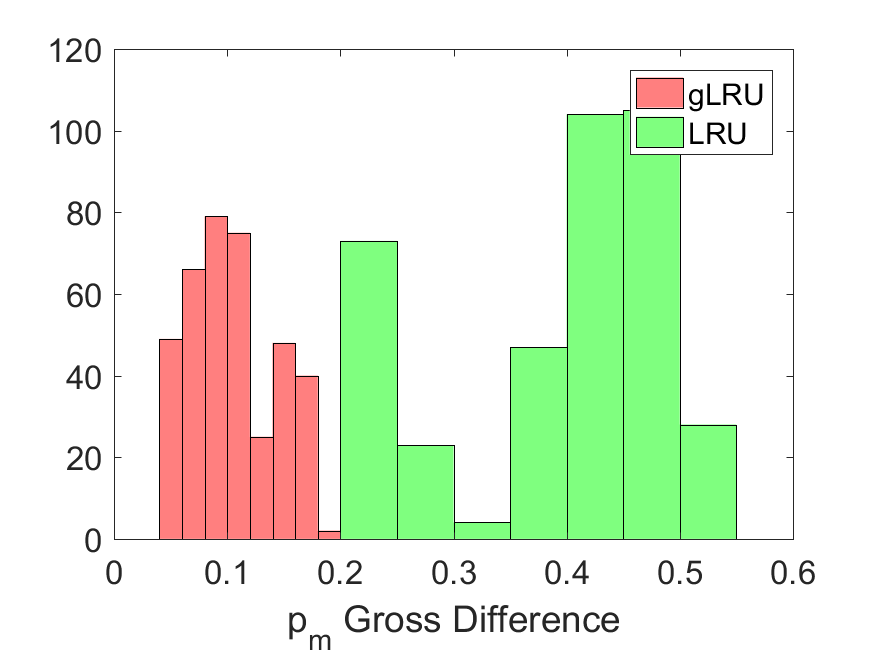}\caption{Cache Miss Rate}
\end{subfigure}\quad
\begin{subfigure}[t]{0.3\textwidth}
\includegraphics[width = 1 \textwidth]{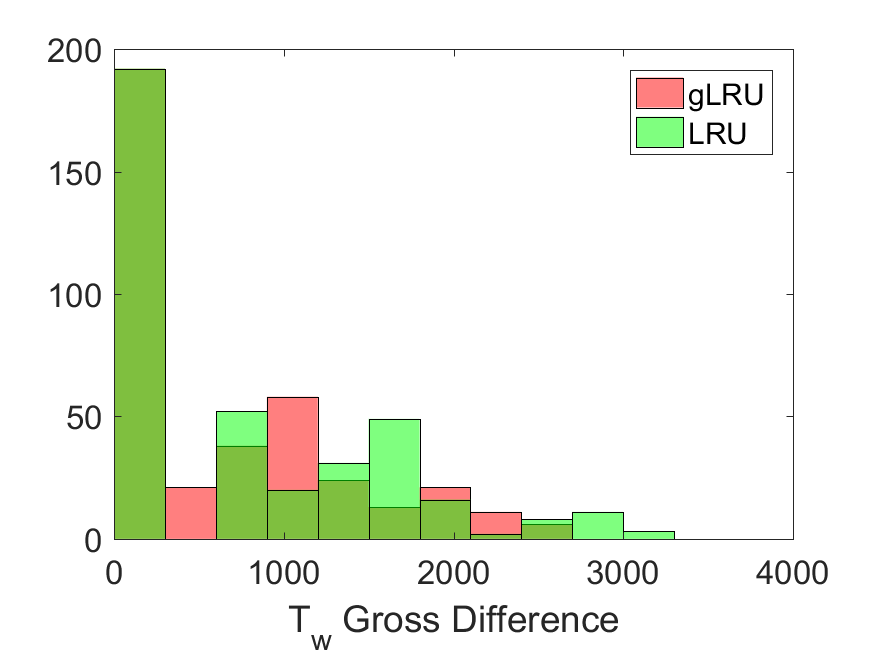}\caption{Download Time}
\end{subfigure}\\

\begin{subfigure}[t]{0.3\textwidth}
\includegraphics[width = 1\textwidth]{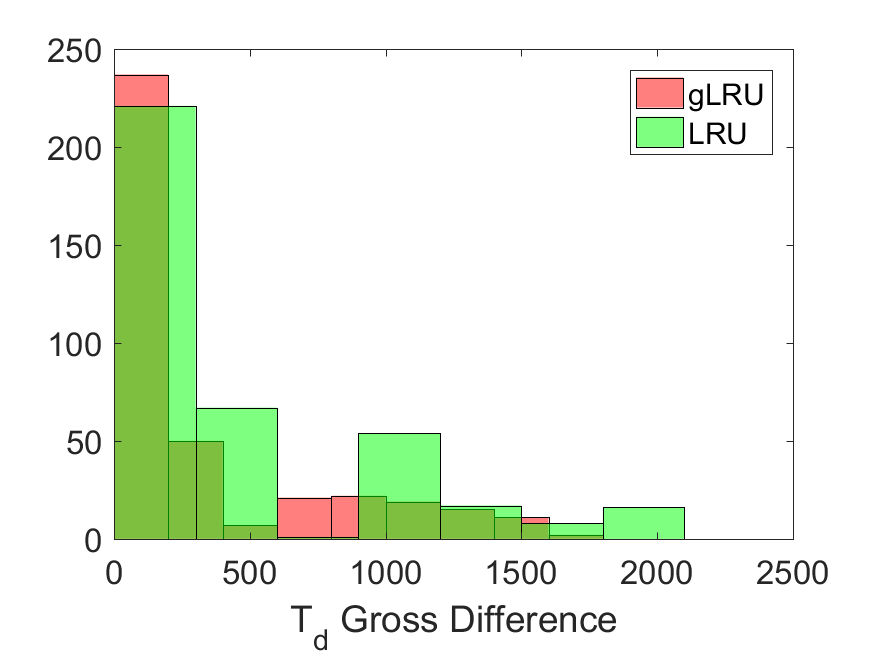}\caption{Delay Time}\label{fig:histtdgr}
\end{subfigure}\quad
\begin{subfigure}[t]{0.3\textwidth}
\includegraphics[width = 1\textwidth]{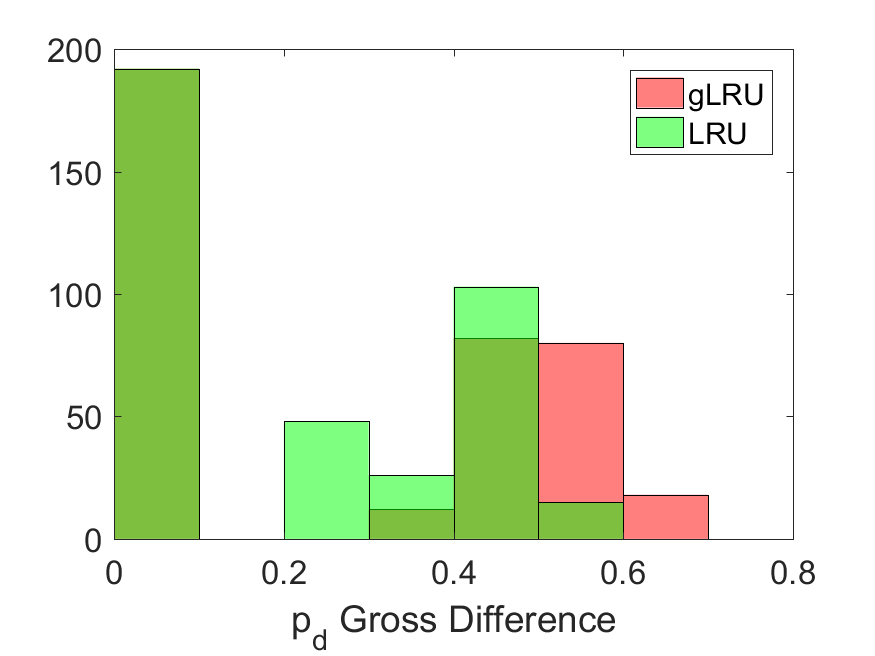}\caption{Proportion Delayed}
\end{subfigure}\caption{Histograms of Relative Performance Improvement of gLRU over LRU.}
\label{fig:histcomp}
\end{figure*}

Histograms for the same performance metrics as considered in \ref{sec:videoStream} are presented in Figure \ref{fig:histcomp}.
In order to compare the case of positive correlation with the case of negative correlation we simply subtract the performance metric of interest obtained under positive correlation from the respective result under negative correlation.
Each subfigure contains two histograms, one for the system under gLRU (red) and one for the system under LRU (green) and provide a more granular view of the data than that was shown in Section \ref{sec:dependence}.
